\documentclass[]{aa}
\usepackage{graphicx}
\usepackage{txfonts}
\usepackage{subfigure}
\usepackage{natbib}
\usepackage{hyperref}
\usepackage{float}
\begin{document}

\title{Characterising the high-mass star forming filament G351.776--0.527 with \textit{Herschel}\thanks{\textit{Herschel} is an ESA space observatory with science instruments provided by European-led Principal Investigator consortia and with important participation from NASA.} and APEX\thanks{This publication is based on data acquired with the Atacama Pathfinder Experiment (APEX). APEX is a collaboration between the Max-Planck-Institut f\"ur Radioastronomie, the European Southern Observatory, and the Onsala Space Observatory.}   dust continuum and  gas observations}

\author{S. Leurini
        \inst{1}
    \and E. Schisano\inst{2,3}
    \and T. Pillai\inst{4,5}
    \and A. Giannetti\inst{3,4}
    \and J. Urquhart\inst{6,4}
    \and T. Csengeri\inst{4}
    \and S. Casu\inst{1}
    \and M. Cunningham\inst{7}
    \and D. Elia\inst{2}
    \and P. A. Jones\inst{7}
    \and C. K\"onig\inst{4}
    \and S. Molinari\inst{2}    
    \and T. Stanke\inst{8}
    \and L. Testi\inst{8,9,10}
    \and F. Wyrowski\inst{4}
    \and K. M. Menten\inst{4}
}

\institute{INAF - Osservatorio Astronomico di Cagliari, Via della Scienza 5, I-09047 Selargius (CA), Italy\\
\email{silvia.leurini@inaf.it}
\and Istituto di Astrofisica e Planetologia Spaziali - INAF, Via Fosso del Cavaliere 100, I-00133 Roma, Italy
\and INAF - Istituto di Radioastronomia, and Italian ALMA Regional Centre, via P. Gobetti 101, I-40129, Bologna, Italy
\and Max-Planck-Institut f\"ur Radioastronomie, Auf dem H\"ugel 69, 53121 Bonn, Germany
\and Institute for Astrophysical Research, Boston University, 725 Commonwealth Ave, Boston, MA, 02215, USA
\and Centre for Astrophysics and Planetary Science, University of Kent, Canterbury CT2 7NH, UK
\and School of Physics, University of New South Wales, Sydney, NSW 2052, Australia
\and European Southern Observatory, Karl-Schwarzschild-Str. 2, 85748, Garching bei M\"unchen, Germany
I               \and INAF-Osservatorio Astrofisico di Arcetri, L.go E. Fermi 5, 50125 Firenze, Italy
                \and Excellence Cluster Universe, Boltzmannstr. 2, 85748, Garching bei M\"unchen, Germany
}
\date{\today}

\abstract
    {G351.776-0.527 is among the most massive, closest,  and youngest filaments in the inner Galactic plane and therefore it is an ideal laboratory to study the kinematics of dense gas and mass replenishment on a large scale. In this paper, we present far-infrared (FIR) and submillimetre wavelength continuum observations  combined with spectroscopic C$^{18}$O\,(2--1)
      data of the entire region to study its temperature, mass distribution, and kinematics. The structure is composed of a main elongated region with an aspect ratio of $\sim 23$, which  is associated with a  network of filamentary structures. The main filament has a remarkably constant width of 0.2\,pc. The total mass of the network (including the main filament) is  $\geq 2600$\,M$_\odot$, while we estimate a mass of $\sim 2000$\,M$_\odot$ for the main structure.  Therefore, the network   harbours a large reservoir  of  gas  and dust that could still be accreted onto the main structure. From the analysis of the gas kinematics, we detect two velocity components in the northern part of the main filament. The data also reveal velocity oscillations in C$^{18}$O
      along the spine in the main filament and in at least  one of the branches. Considering the region as a single structure, we find that it is globally close to virial equilibrium indicating that the entire structure is  approximately in a stable state.}
\titlerunning{High-mass star forming filament G351.776--0.527}
\authorrunning{Leurini et al.}
\maketitle
 
%

\date{\today}




\section{Introduction}\label{sec_intro}

In recent years increasing evidence has been collected that 
high-mass  star  formation  is  tightly  linked  to
the formation of massive clumps and massive clusters.
Observations in molecular tracers  \citep[e.g.][]{2010A&A...520A..49S,2013A&A...555A.112P,2014A&A...561A..83P,2017A&A...602L...2H} have revealed global collapse in several massive clouds and have suggested that clumps   build up their mass as a result of the supersonic global collapse of their surrounding cloud. For instance, \citet{2014A&A...561A..83P} suggested that organised velocity gradients in the filaments of SDC13 are the result of large-scale longitudinal collapse and could generate kinematic support against fragmentation, helping the formation of super-Jeans cores. Also in the case of DR21, \citet{2010A&A...520A..49S} detected signs of global collapse of the filament and suggested that its gas content is continuously replenished via sub-filaments.
In the case of  intermediate-mass clouds, \citet{2013MNRAS.432.1424K}
identified gas motions onto the filament along defined structures  and from the filament onto the central cluster in the Serpens region. 
Recently, \citet{2017arXiv170600118M} proposed an evolutionary scenario in which global collapse of massive clouds and hubs generates gas
flow streams.
In this picture, such flows would efficiently  concentrate matter on small scales, helping to increase the mass of  low-mass prestellar seeds  present in the cloud.
 With time these seeds eventually become  high-mass protostars. Numerical simulations also support this scenario \citep[e.g.][]{2017MNRAS.467.1313V}. Observational evidence of gas streams inflowing on small scales on protostars  is reported by \citet{2011ApJ...740L...5C}.

 Detailed studies of massive filaments and of their internal structure and  dynamics  are  crucial to better understand the link between high-mass star formation and  cloud and cluster formation. Recently, unbiased surveys of the Galactic plane at FIR and submillimetre wavelengths, namely the APEX Telescope Large Area Survey of the Galaxy (ATLASGAL) and the Herschel infrared Galactic Plane Survey (Hi-GAL) \citep[][respectively]{2009A&A...504..415S,2010A&A...518L.100M},    have allowed for the compilation of catalogues of candidate filaments over the entire plane \citep[e.g.][]{2016A&A...591A...5L,2016ApJS..226....9W,Schisano18}. These studies have shown that massive filaments are in general a few kiloparsec distant \citep[e.g.][]{2016A&A...591A...5L,Schisano18}. Therefore high angular resolution observations are needed to resolve linear scales below 0.1--0.2\,pc  to resolve the kinematics towards individual cores. These catalogues are ideal to select  samples of filaments within a few kiloparsecs  from the Sun that are suitable for a detailed investigation of the link between massive star formation and cluster formation.

\begin{figure*}[!htb]
\centering
        \includegraphics[width=0.9\textwidth]{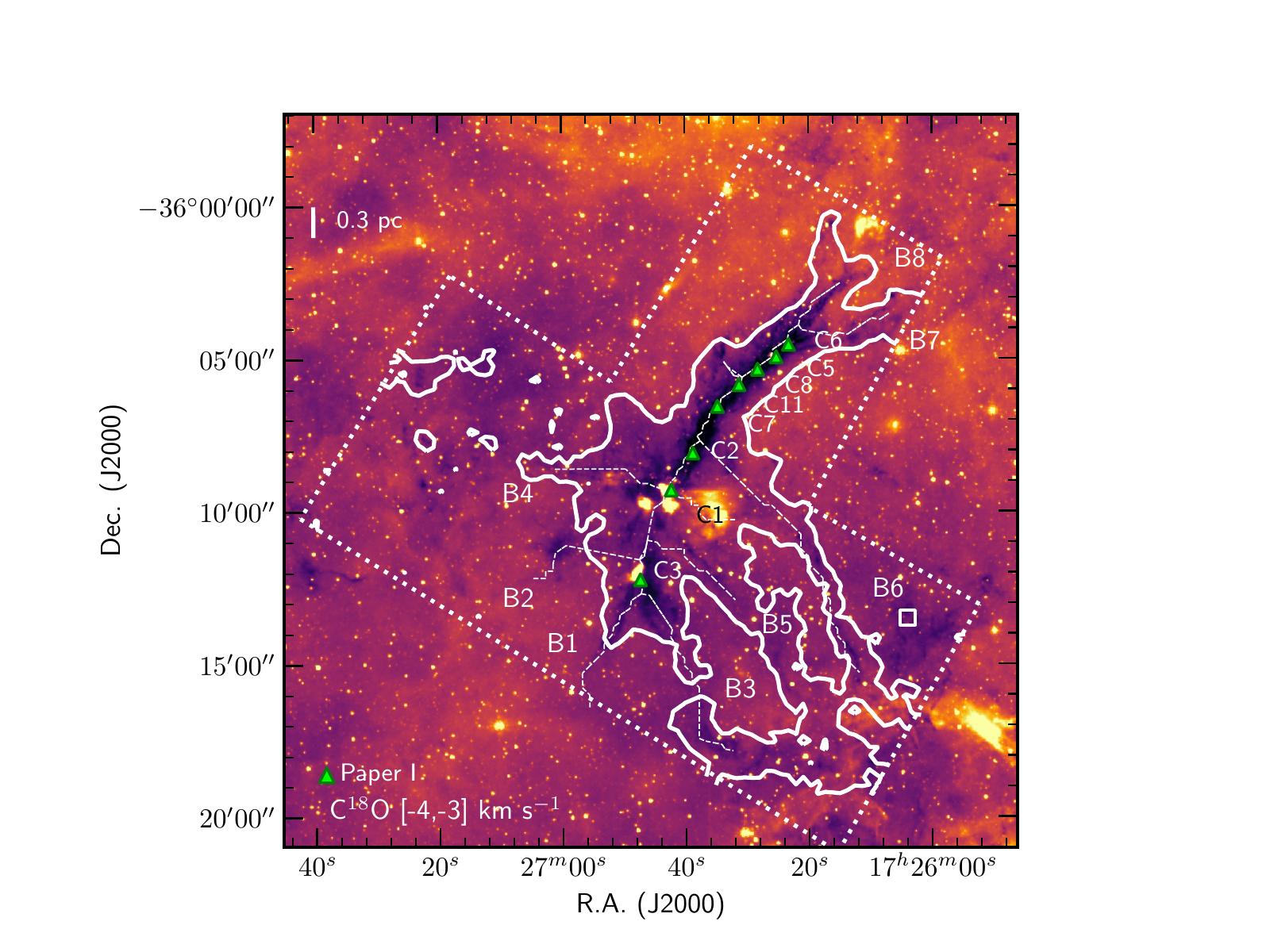}
    \caption{{\it Spitzer} 8\,$\mu$m image of the molecular complex G351. The white contour is 5\% of   the C$^{18}$O\,(2--1) peak intensity (17.6 K\,km\,s$^{-1}$) integrated  in the velocity range $-3.5\pm0.5$\,km\,s$^{-1}$. 
    The green triangles denote the ATLASGAL clumps identified in Paper\,I (see Table\,\ref{tab1}, labelled as C1, C2...). The 
 skeleton of G351 and that of the branches is shown by the white dashed line. The white dotted lines indicate the region observed in CO isotopologues with APEX. The B1--B8 labels indicate the eight branches identified in the $N_{\rm{H_2}}$ column density map (see Sect.\,\ref{sec_dust}). The white square on the south-west of G351 indicates the position of MSX IRDC G351.64--00.46 \citep{2006ApJ...639..227S}.} 
    \label{fig1}
\end{figure*}

 In this paper, we investigate the internal structure of the close-by massive infrared dark cloud (IRDC) G351.776--0.527  (hereafter G351) and that of the network of filaments surrounding this cloud. The study deals with the large-scale filamentary environment of the source, while our previous works \citep{2008A&A...485..167L,2009A&A...507.1443L,2011A&A...530A..12L,2013A&A...554A..35L,2014A&A...564L..11L}  focussed on the star forming region IRAS\,17233--3606 (Clump-1, see Table\,\ref{tab1} and Fig.\,\ref{fig1}).
 Given the fortuitous proximity of the source ($D=0.7-1$\,kpc, \citealt{2011A&A...533A..85L},  hereafter Paper\,I; \citealt{2015A&A...579A..91W}, and discussion therein), G351 is an ideal target to study  a massive filamentary network and its dynamics in detail.
The paper is structured as follows.  In Sect.\,\ref{sec_sou}, we discuss the  properties of the source and explain the uniqueness of this laboratory.  In Sect.\,\ref{sec_obs}, we
describe the spectroscopic and continuum observations used in this paper to further characterise G351. 
In Sect.\,\ref{sec_lines}, we study
the main properties of G351 and derive its temperature and mass distribution. Section\,\ref{sec_vel} is devoted  to the 
analysis of the gas velocity field.
In this section, we demonstrate the velocity coherence of the whole filamentary network, and study the velocity structure of the diffuse/dense gas associated with the source. Finally we discuss the dynamical state of G351 in Sect.\,\ref{sec_virial}.

\section{G351.77--0.51: A site of massive star formation}\label{sec_sou}

Thanks to  Galactic plane surveys at different wavelengths and tracers performed in recent years \citep[e.g.][]{2006ApJS..163..145J,2009A&A...504..415S,2010A&A...518L.100M,2013PASA...30...44B,2015ApJ...812....6B,2017A&A...601A.124S}, catalogues of interstellar filaments are now available \citep[e.g.][]{2014A&A...568A..73R,2016A&A...591A...5L,2016ApJS..226....9W,Schisano18}. These catalogues have shown that massive filaments are on average a few kpc in distance from the Sun:  \citet{Schisano18} found for example that the average mass of Hi-GAL filaments within 3\,kpc varies between 200\,M$_\odot$ and   800\,M$_\odot$ with a binning of 200\,pc. Candidates from ATLASGAL are in general more massive given the lower sensitivity of the survey, which is not sensitive to low-mass features, and have an average mass of 470\,M$_\odot$--1700\,M$_\odot$ within 3\,kpc (in binning of 1\,kpc given the low source density,  \citealt{2016A&A...591A...5L}).

Despite the uncertainty in its distance, G351 is the most massive ($\sim 1300$\,M$_\odot$ for $D=0.7$\,kpc, \citealt{2016A&A...591A...5L}) and closest filament identified in the ATLASGAL
survey of the inner Galactic plane \citep{2016A&A...591A...5L}. Compared to  close-by massive filaments such as the Orion Integral Filament and NGC\,6334 with  masses of $\sim 10\,000-15\,000$\,M$_\odot$ \citep{1987ApJ...312L..45B,2017A&A...602A..77T}  at $\sim0.4$\,kpc and $1.7$\,kpc, respectively \citep{2007A&A...474..515M,2012A&A...538A.142R}, G351 is a younger, less complex object that has only two sites of active massive star formation (Clump-1 and -2, see Table\,\ref{tab1} and Fig.\,\ref{fig1} and discussion below). 
The structure is composed by a main elongated region with an aspect ratio of $\sim 23$ (see Sect.\,\ref{sec_g351}), which appears nested in a remarkable network of filamentary structures over its whole length (see Fig.\,\ref{fig1}). In this work we adopt the term branch to refer to these sub-filaments following the naming convention of \citet{2014ApJ...791...27S}.
The main filamentary body has been identified as coherent in velocity at $\varv_{LSR}\approx\,-3$\,km\,s$^{-1}$  with a velocity of shift $>1.6$\,km\,s$^{-1}$ between the northern and southern ends (Paper\,I).
Before this work there was no information available on the branches that appear to be connected to the main structure.
These structures are  seen in extinction at 8\,$\mu$m and 24\,$\mu$m, and  they are visible in emission at longer  wavelengths in Hi-GAL images  for $\lambda\geq\,$250\,$\mu$m  (see Fig.\,\ref{fig_higal}) and, as a consequence, in the \textit{Herschel} derived column density map \citep{2014ApJ...791...27S}.

In Paper\,I, we analysed the dust continuum emission at 870\,$\mu$m associated with  G351 and identified  nine dust clumps (see Table\,\ref{tab1} and Fig.\,\ref{fig1}) with masses ranging between  $10\,\rm{M_{\odot}}-430$\,M$_{\odot}$ depending on the temperature (see Table\,4 in Paper\,I). 
These  clumps are confirmed in the ATLASGAL  Gauss clump catalogue \citep{2014A&A...565A..75C} with the exception of Clump-4, which is probably an artefact introduced by CLUMPFIND\footnote{http://www.ifa.hawaii.edu/users/jpw/clumpfind.shtml} in our previous analysis.
The same clumps are also confirmed in the Hi-GAL photometric catalogue \citep{2016A&A...591A.149M}; roughly another 20 sources are identified in the filamentary network in the region shown in Fig.\,\ref{fig1} in at least the  160\,$\mu$m, 250\,$\mu$m, and 350\,$\mu$m Hi-GAL  catalogues. We will provide a detailed analysis of the compact sources in a forthcoming paper; in this work we 
only focus on the massive star formation taking place in G351.

In Paper\,I, we showed that the majority of the clumps in G351 (Clump-1 for a dust temperature, $T_{\rm{dust}}$, of 35\,K, Clump-2 and Clump-5 for $T_{\rm{dust}}=25$\,K, Clump-6, Clump-7, Clump-8, and Clump-11 for 10\,K; see Sect.\,\ref{sec_dust} for the temperature distribution in the source and Table\,4 in Paper\,I for the masses of the clumps) exceed the threshold for massive star formation determined by  \citet{2010ApJ...723L...7K}. At least two clumps (Clump-1 and Clump-3)   host active massive star formation. Clump-1 is
the brightest submillimetre condensation saturated in Hi-GAL maps (see Fig\,\ref{fig_higal}); this clump corresponds to the FIR source IRAS\,17233--3606, and
hosts an H{\sc ii} region \citep{1993AJ....105.1495H}, a   hot molecular core  \citep{2008A&A...485..167L,2011A&A...530A..12L}, and a cluster of radio continuum sources \citep{2008AJ....136.1455Z}. Within 10\arcsec\, of Clump-3 , an H{\sc ii} region has been identified in data taken with the Wide-field Infrared Survey Explorer (WISE) \citep{2014ApJS..212....1A}. Apart from Clump-1, there is little evidence for recently formed high-mass stars in the region. 
In Table\,\ref{tab1}, we classify the clumps in G351 according to their far- and mid-IR fluxes as follows: 70\,$\mu$m weak, when not detected or with weak emission at 70\,$\mu$m in Hi-GAL (Clump-7 and Clump-11, upper limits $F_{70\mu\rm{m}}< 2.5$\,Jy, and $< 2.0$\,Jy, respectively); mid-IR weak if detected in PACS-70\,$\mu$m, but with a 21--24\,$\mu$m flux below 2.6\,Jy; and mid-IR bright when the  21--24\,$\mu$m flux is larger than 2.6\,Jy, following the procedure and the classification scheme adopted by \citet{2017A&A...599A.139K} and \citet{2018MNRAS.473.1059U} for the ATLASGAL catalogues \citep{2013A&A...549A..45C,2014A&A...568A..41U,2014A&A...565A..75C}.
Further evidence of star formation activity is provided by other tracers at several other  positions along G351.
Four clumps (Clump-1, Clump-2, Clump-3, and Clump-5)   show signs  of jets and outflows detected in {\it Spitzer} IRAC band 3 (see Paper\,I).

The vicinity of G351,  its very early evolutionary phase with only two active sites of massive star formation, and the remarkable  filamentary network of branches provide a great opportunity to study the kinematics of dense gas,  investigate the mass replenishment from the surrounding environment into sites of massive star formation, and assess the potential for further star formation. In the following discussion, we  adopt a
distance of 1\,kpc for simplicity.

\begin{table*}
\centering
\caption{Clumps in G351}\label{tab1}
\begin{tabular}{lccc}
\hline
\multicolumn{1}{c}{Clump Id} &\multicolumn{1}{c}{R.A.}&\multicolumn{1}{c}{Dec.}&\multicolumn{1}{c}{Classification}\\
\multicolumn{1}{c}{} &\multicolumn{1}{c}{(J2000)}&\multicolumn{1}{c}{(J2000)}\\
\hline
Clump-1  & 17:26:42.30  &   $-$36:09:18.23&Mid-IR bright/H{\sc ii} region\\
Clump-2  & 17:26:38.79  &   $-$36:08:05.53&Mid-IR weak\\
Clump-3  & 17:26:47.33  &   $-$36:12:14.37&Mid-IR bright/H{\sc ii} region\\
Clump-5  & 17:26:25.25  &   $-$36:04:57.03&Mid-IR weak\\
Clump-6  & 17:26:23.25  &   $-$36:04:32.74&Mid-IR weak\\
Clump-7  & 17:26:34.78  &   $-$36:06:34.22&70\,$\mu$m-IR weak\\
Clump-8  & 17:26:28.26  &   $-$36:05:21.34&Mid-IR weak\\
Clump-11 & 17:26:31.27  &   $-$36:05:51.70&70\,$\mu$m-IR weak\\
\hline
\end{tabular} 
\end{table*}

\section{Observations}\label{sec_obs}
In this paper we present spectroscopic and continuum  observations of the molecular complex G351 at (sub)millimetre  wavelengths with the APEX telescope. 
We complement these data  with the \textit{Herschel} maps from the Hi-GAL survey \citep{2010A&A...518L.100M} extracting a region of approximately $0.5^\circ \times 0.5^\circ$ centred on G351 from the flux calibrated maps of the data release version 1 \citep{2016A&A...591A.149M}. 
{\it Spitzer} 8\,$\mu$m and  24\,$\mu$m data are also used for comparison.
A summary of the datasets used in this study is presented in Table\,\ref{tab2}.

\begin{table}
\centering
\caption{Summary of the datasets available at different wavelengths}\label{tab2}
\begin{tabular}{lcc}
\hline
&Angular resolution&References\\
\hline
\multicolumn{3}{c}{Continuum observations}\\
\it{Spitzer}-8\,$\mu$m&1\farcs2&1\\
\it{Spitzer}-24\,$\mu$m&5\farcs75&2\\
\it{Herschel}-70\,$\mu$m&10\arcsec&3\\
\it{Herschel}-160\,$\mu$m&12\arcsec&3\\
\it{Herschel}-250\,$\mu$m&18\arcsec&3\\
\it{Herschel}-350\,$\mu$m&25\arcsec&3\\
\it{Herschel}-500\,$\mu$m&36\arcsec&3\\
SABOCA-350\,$\mu$m&7\farcs4&4\\
LABOCA-870\,$\mu$m&19\farcs2&5\\
\hline
\multicolumn{3}{c}{Spectroscopic observations}\\
APEX-C$^{18}$O\,(2--1)&30\farcs2&4\\
\hline
\end{tabular} 
\tablebib{(1) \citet{2003PASP..115..953B,2009PASP..121..213C}; (2) \citet{2009PASP..121...76C}; (3) \citet{2010A&A...518L.100M}; (4) This study; (5) \citet{2009A&A...504..415S} and Paper\,I
}
\end{table}

\subsection{Spectroscopic observations}\label{sec_spectra}
The region indicated by white dotted lines in Fig.\,\ref{fig1} was mapped in the $^{13}$CO and C$^{18}$O\,(2--1) transitions with the APEX telescope between August and November 2014. The APEX-1 facility receiver was tuned to a frequency of 218.5\,GHz in lower side band (LSB) to observe simultaneously the $^{13}$CO\,(2--1) and C$^{18}$O\,(2--1) transitions with a velocity resolution of  0.1\,km\,s$^{-1}$. The region was covered with three different on-the-fly maps centred on $\alpha({\rm J2000})$ = 17$^h$26$^m$36\farcs47,
$\delta({\rm J2000})$ = --36$\degr$07$\arcmin$43\farcs00, $\alpha({\rm J2000})$ = 17$^h$26$^m$20\farcs00,
$\delta({\rm J2000})$ = --36$\degr$15$\arcmin$15\farcs30, and  $\alpha({\rm J2000})$ = 17$^h$27$^m$14\farcs84,
$\delta({\rm J2000})$ = --36$\degr$08$\arcmin$07\farcs70, respectively. This results in a non-uniform noise in the map due to different integration times for the three coverages. For all observations, the position 
$\alpha({\rm J2000})$ = 17$^h$26$^m$36\farcs47 $\delta({\rm J2000})$ = --38$\degr$07$\arcmin$43\farcs09 was used for sky subtraction after verifying that no $^{13}$CO\,(2--1) emission was detected close to the ambient velocity of the G351 complex ($\sim -3$\,km\,s$^{-1}$). Pointing and focus were regularly checked on Mars and on nearby stars (NGC6302, IRAS15194--5115) of the APEX line pointing catalogue\footnote{\url{http://www.apex-telescope.org/observing/pointing/lpoint/}}. For relative calibration,  the hot core IRAS\,17233--3606 \citep{2011A&A...530A..12L} was  regularly observed in the same frequency set-up used for the maps; variations in the $^{13}$CO\,(2--1) line intensity  were found to be less than 15\%.  Data were converted into $T_{\rm MB}$ units assuming a forward efficiency of 0.95 and a beam efficiency of 0.75. Final data cubes were constructed  using the CLASS\footnote{The CLASS program is part of the GILDAS software package \url{http://www.iram.fr/IRAMFR/GILDAS}} gridding routine XY\_map, which convolves the    data with a Gaussian kernel of one-third of the telescope beam,   yielding a  final  angular resolution slightly coarser than the original beam size. The final spatial resolution of the maps is 30\farcs2. The average rms in the maps is 0.4\,K per velocity channel (0.1\,km\,s$^{-1}$) with a spread between 0.2 and 0.5\,K owing to non-uniform integration time. The region of the map with the higher signal-to-noise ratio covers the main filament and the branches detected in absorption in the mid-IR.

\subsection{SABOCA continuum observations}\label{sec_saboca}

G351 was observed with the APEX telescope in the continuum emission at 350 $\mu$m with 
the Submillimeter APEX Bolometer Camera \citep[SABOCA; ][]{2010Msngr.139...20S}. The observations were performed on 2010 May 7 and 8. 
The total mapped area  is $\sim 3\farcm5 \times 11\arcmin$ orientated along the main filament G351 and covers a smaller area than that of the C$^{18}$O data.
The pointing and focus were checked on SABOCA secondary calibrators sources (G45.1, IRAS\,16293) and planets (Uranus and Neptune, also used as primary calibrators)\footnote{\url{http://www.apex-telescope.org/bolometer/saboca/calibration/\#calibrators}};  the bright hot core IRAS\,17233--3606  in the G351 complex was also used as pointing source close to the target. Skydips (fast scans in elevation at constant azimuthal angle) were performed to estimate the atmospheric opacity ($\tau_{\rm{zenith}}\sim 1.2$). The weather conditions at the time of the observations were good with  precipitable water vapour levels of 0.3--0.5~mm. The data were reduced with the BOA software \citep{2012SPIE.8452E..1TS}. We note that the SABOCA data  are affected by spatial filtering of extended structures due to the standard data reduction, which relies on the process of correlated noise removal \citep[see for example][for the LABOCA array]{2011A&A...527A.145B}. 
The  resolution of the data is 7\farcs4 with a noise level of 0.9\,Jy\,beam$^{-1}$.

\section{Dust and molecular environment of G351}\label{sec_lines}
In this section, we discuss the dust continuum emission of G351 (Sect.~\ref{sec_dust}), and
identify a main body filamentary structure based on the H$_2$ column density distribution (Sect.~\ref{sec_g351}). We also investigate the molecular environment of the cloud (Sect.~\ref{sec_co}), and derive the mass distribution of the entire filamentary network (Sect.~\ref{sec_mass}) by means of  the dust and molecular data.

\subsection{Dust emission}\label{sec_dust}

\begin{figure*}[!htb]
\includegraphics[width=0.99\textwidth]{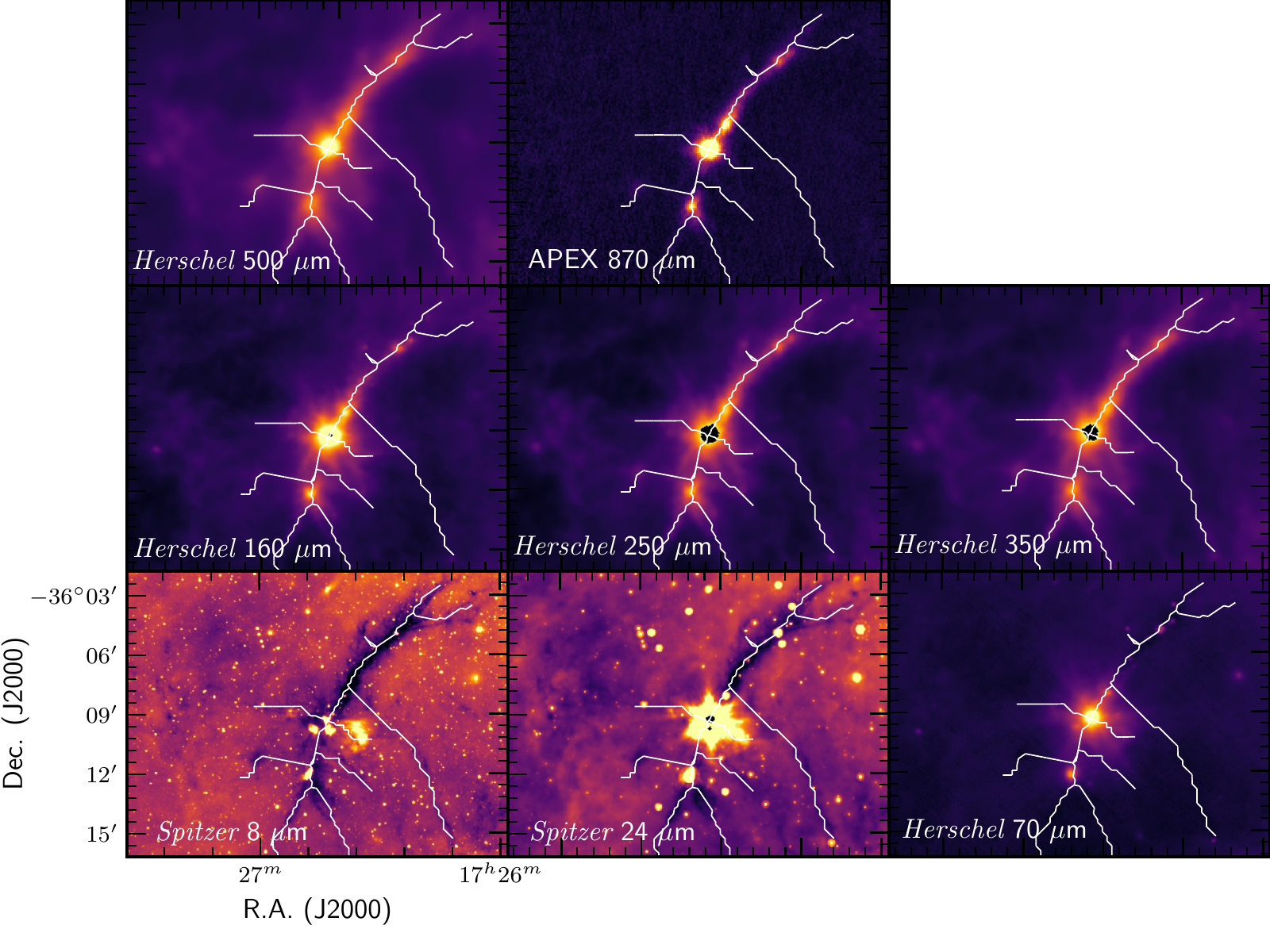}
\caption{Continuum images of G351 at different wavelengths from 8\,$\mu$m with {\it Spitzer} (bottom left panel) to 870\,$\mu$m with  ATLASGAL-LABOCA. The LABOCA  data are affected by filtering of large-scale emission \citep[e.g.][]{2009A&A...504..415S}. The solid white line in each panel denotes the 
 skeleton of G351 and of the branches.}
\label{fig_higal}
\end{figure*}

In Fig.\,\ref{fig_higal} we show the continuum emission of the region from 8\,$\mu$m to 870\,$\mu$m using complementary \textit{Spitzer} and ATLASGAL data. 
The high resolution of mid-IR \textit{Spitzer} data \citep{2003PASP..115..953B,2009PASP..121..213C,2009PASP..121...76C} allow us to distinguish  fine details of the filamentary network. In particular, we note several branches in the north giving this region the appearance of a ripped fan. 
This configuration is seen in several IRDCs and it is expected to appear in models of filaments in global collapse with active environmental accretion onto their structure \citep{2013ApJ...769..115H}. 
The whole filamentary system (including G351 and the network of branches surrounding it)  is seen in absorption at 8\,$\mu$m and 24\,$\mu$m. The structure starts to become detected in emission at longer wavelengths. For example, at 70\,$\mu$m the northern part of G351 is still in absorption against the  Galactic background. However, the 70\,$\mu$m dark region is narrower than at shorter wavelengths and it corresponds to the inner part in absorption at 8\,$\mu$m and 24\,$\mu$m. In the other Hi-GAL maps, G351 and the branches are detected in emission.

The 160\,$\mu$m to 500\,$\mu$m Hi-GAL data were used to derive a map of the dust temperature and  ${\rm{H_2}}$ column density, $N_{\rm{H_2}}$. We  smoothed all data to the beam size of the   500\,$\mu$m map following a standard approach in the analysis of \textit{Herschel} data \citep[e.g.][]{2013ApJ...772...45E,2012A&A...540L..11S,2017MNRAS.471..100E}.  The derived maps have a final resolution of $36\arcsec$, which matches very closely that of the spectroscopic data analysed in this paper.
  To determine the column density and dust temperature,  we used  a two emission component model \citep{Schisano18}, in which the fluxes in each pixel at each \textit{Herschel} band are split between a filament and background component as discussed by \citet{2010A&A...518L..98P}. The background emission is estimated through a linear interpolation over the area of G351 of the fluxes measured around its perimeter. The interpolation direction is aligned perpendicularly to the branches (see Sect.\,\ref{sec_lines}).
The fluxes are then fit with a modified black-body model, deriving an independent temperature and column density for each
component. In both cases, the dust temperature is constrained to be in the range 5\,K -- 50\,K, while no constraint was adopted for the column density estimate.   The fluxes were fitted using a modified black-body function as in   \citet{2013ApJ...772...45E} and the best fit is found
with a $\chi^2$ analysis.
  We assumed a dust opacity law $\kappa_0\,(\nu/\nu_0)^\beta$ with $\beta=2$, $\nu_0=1250$\,GHz, a dust-to-mass ratio of 100, and $\kappa_0=0.1$\,cm$^2$\,g$^{-1}$ \citep{1983QJRAS..24..267H} 
to preserve  compatibility with previous works based on other Hi-GAL studies \citep[e.g. ][]{
2014ApJ...791...27S,2017MNRAS.471..100E}. 
\textit{Herschel} data are saturated at 250\,$\mu$m and 350\,$\mu$m in a region of  $25\arcsec$ radius centred on Clump-1. Saturated pixels are also detected at 160\,$\mu$m and 500\,$\mu$m around Clump-1 peak position caused by its high brightness. To avoid possible non-linear effects, we masked out a region of radius  $\sim 30\arcsec$ in all \textit{Herschel} bands before computing $T_{\rm{dust}}$ and $N_{\rm{H_2}}$. Nevertheless, the  region surrounding Clump-1 peak's position is affected by larger uncertainties in both quantities. 
The resulting $T_{\rm{dust}}$ and $N_{\rm{H_2}}$ distributions of the network are shown in Fig.\,\ref{fig4}, while the corresponding maps for the background are presented in Appendix\,\ref{app_fig} (Fig.\,\ref{bkgrd}).

The main filamentary body of G351 is the denser region of the cloud, which has values exceeding $\geq 3\times10^{22}$\,cm$^{-2}$; the branches, on the other hand, have lower column densities with a maximum value of $\sim 1.5\times 10^{22}$\,cm$^{-2}$ showing a main  difference between the main body and the lower surface brightness shallow surrounding filamentary network.    The average  column density along the skeleton (see discussion below) is $\sim 8.2 \times 10^{22}$\,cm$^{-2}$ in G351. The dust temperature is distributed differently along the main filament as it is influenced by local star formation.  Indeed, the more active southern part  (see also Paper\,I) is warmer with typical temperatures of 27\,K--40\,K, reaching the highest values in the surroundings of Clump-1, while the more quiescent northern part has typical temperatures of 11\,K--13\,K. The  branches in the network have  temperatures similar to the northern part of G351.

  We  derived the spine of G351 and of the network of branches
  using a twofold approach. For the filamentary network, we made use of the
  column density map in which the branches have the largest contrast with the surrounding
  background. However, the resolution of the column
  density map is $36\arcsec$ and with this approach we lose
  details on the main filament. Therefore, we also computed the
  skeleton of the main body from the isocontours of the 250\,$\mu$m
  continuum emission, which has high signal-to-noise ratio and 18\,\arcsec beam size.
  We adopted the 
   method described by \citet{2014ApJ...791...27S} and
  optimised it for the region.  The adopted algorithm identifies extended
  two-dimensional cylindrical-like features based on the analysis of
  the \textit{Hessian} matrix of the column density map or the intensity map at 250\,$\mu$m.
  We refer to
  \citet{2014ApJ...791...27S} and \citet{Schisano18} for a detailed
  discussion of the method.     
  To overcome  issues related to the saturation
  in the Hi-GAL bands, we linearly interpolated the skeleton in the
  30\,\arcsec in radius masked region. Moreover, we improved the position of the spine by comparing the emission in each of its pixel with that of  the direct neighbours.
  The resulting spine for the entire network
  is shown in Fig.\,\ref{fig1} and \ref{fig_higal}, while that of the main filament derived from the
  250\,$\mu$m data is shown in Figs.\,\ref{saboca} and \ref{saboca_spire}. In the following analysis, we make use of the spine derived from the column density map only in Sect.\,\ref{sec_dv_co} when we analyse the velocity field along one of the branches.

  We identified the main skeleton
  and additional eight branches in the network. The most prominent are on
  the south-west of the main IRDC pointing towards the brightest
  FIR clumps in the filament, i.e. Clump-1, -2, and -3 (see
  Table\,\ref{tab1}); Clump-1 and -3 are also the most evolved
  in the region. All three branches have a dark counterpart at
  8\,$\mu$m. In the following we label these three sub-filaments
  as Branch-3, Branch-5, and Branch-6 (Fig.\,\ref{fig1}). In the
  south-east of G351 there are two other branches (Branch-2 and -4),
  which also seem to point to Clump-3 and -1, respectively. Branch-4
  is also seen in extinction in the mid-IR. Branch-1 is aligned along
  the main axis of the central region.  Finally, two branches
  (Branch-7 and -8) are detected in the most northern extension of
  G351 associated with the fan-like structure seen in the mid-IR.

In the following discussion, we  focus on the large-scale dust distribution. The analysis of the dust condensations detected in Hi-GAL and the protostellar content in  the filamentary network will be presented in a forthcoming paper (Schisano et al. in prep.).

\subsection{Filament main body: G351}\label{sec_g351}
\begin{figure*}[!htb]
\centering
\subfigure[][]{\includegraphics[width=0.45\textwidth]{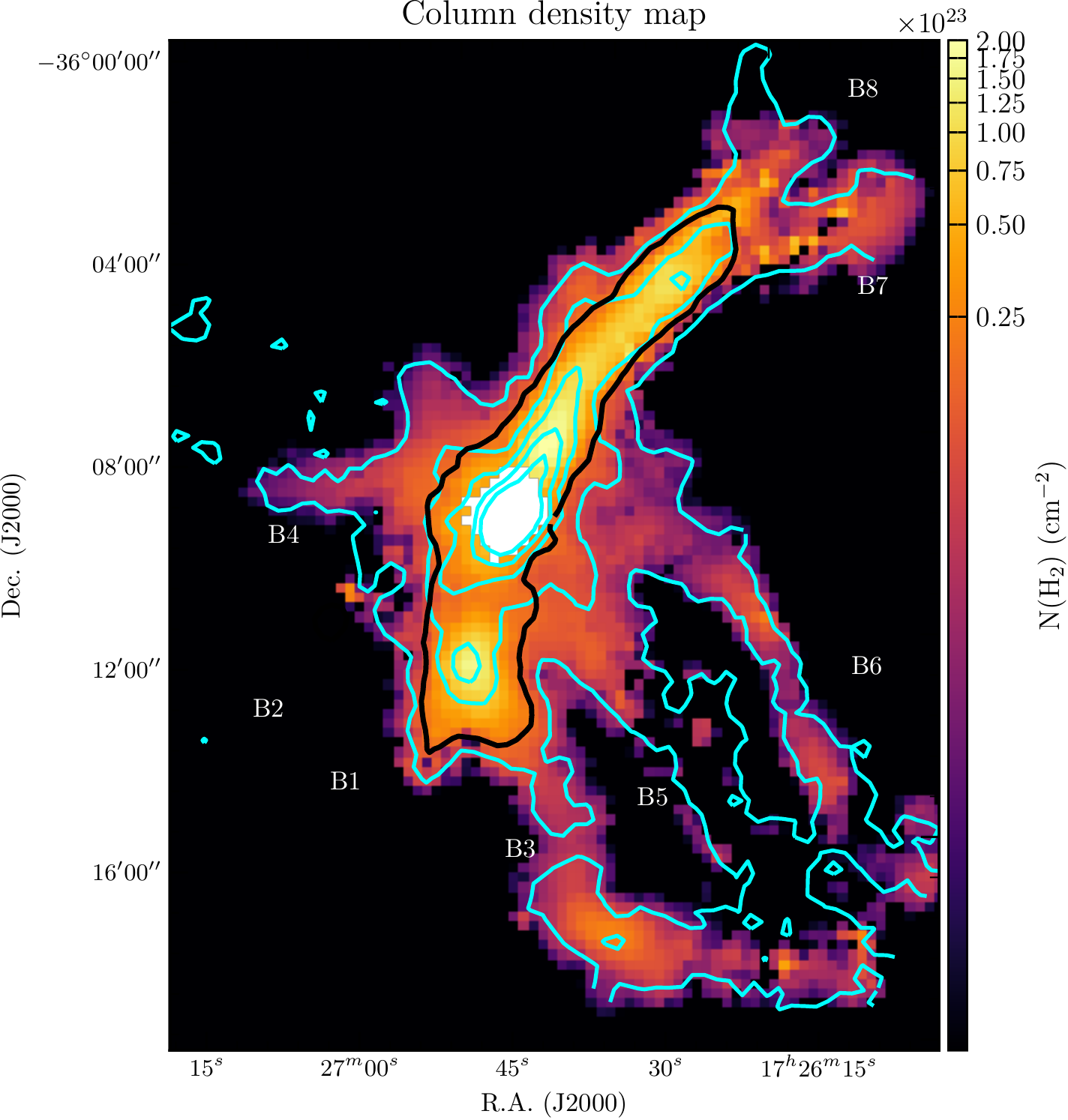}}
\subfigure[]{\includegraphics[width=0.45\textwidth]{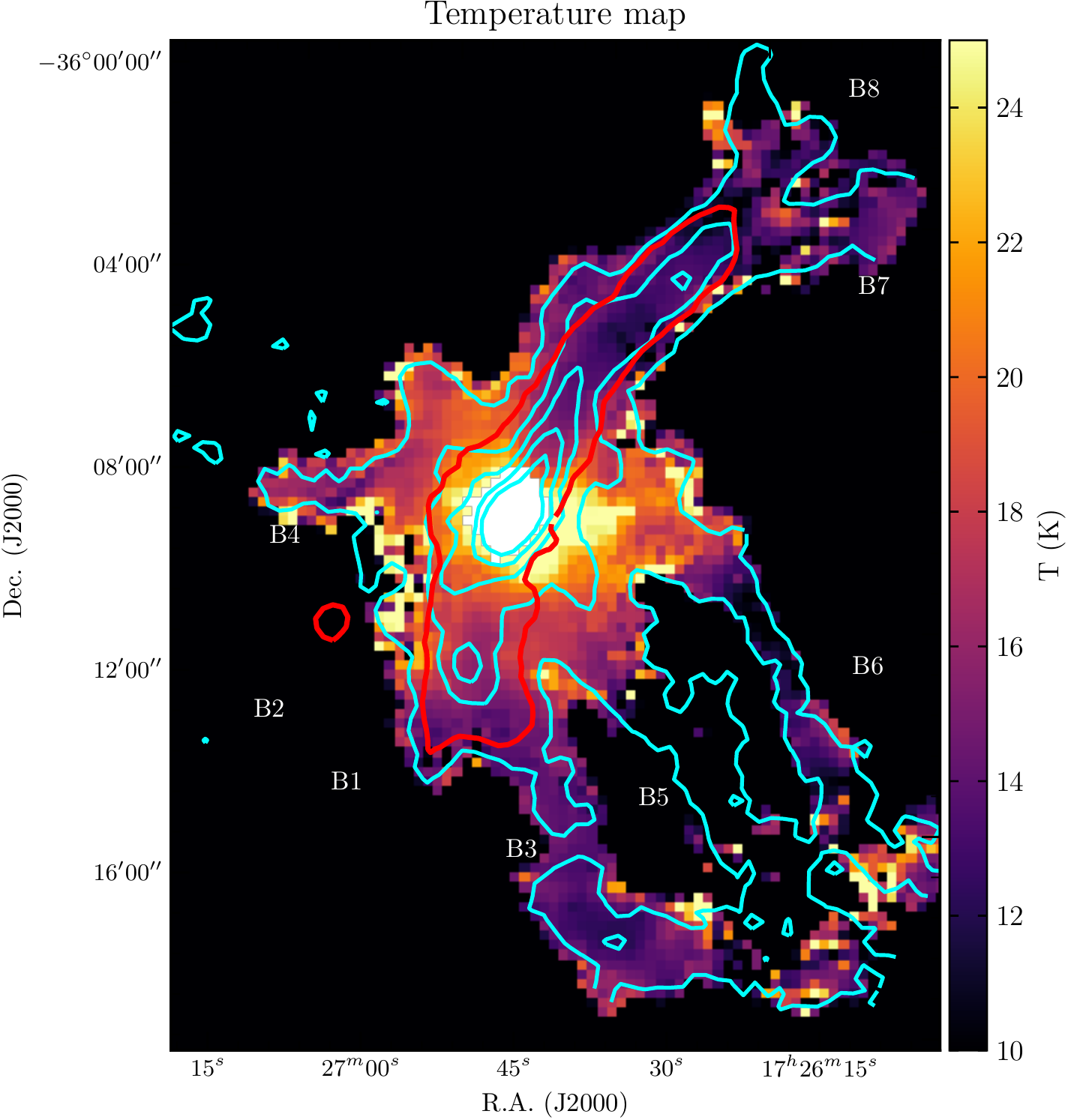}}
    \caption{Distribution of the $N_{\rm{H_2}}$ column density (left panel) and dust temperature (right panel). The cyan contours are the C$^{18}$O integrated emission in the velocity range $-3.5 \pm 0.5$\,km\,s$^{-1}$ from 5\% to 45\% of the peak integrated intensity (17.6\,K\,km\,s$^{-1}$) in steps of 10\%. In panel a, the black solid contour (red in panel b) indicates the region with N$_{\rm H_2}=3\times 10^{22}$\,cm\,$^{-2}$. The B1--B8 labels indicate the eight branches identified in the $N_{\rm{H_2}}$ column density map.}
    \label{fig4}
\end{figure*}

By combining the Hi-GAL data (resolution between $\sim10$\arcsec\, at 70\,$\mu$m
and $\sim36\arcsec$ at 500\,$\mu$m, 0.05\,pc--0.17\,pc, respectively), which sample the larger scale distribution of the dust, with the  SABOCA data (angular resolution $\sim 7\farcs4$, corresponding to 0.04\,pc), which trace the innermost high-density region of the filament, we can construct a detailed view of the dust distribution in G351.

We define as the main body of the region the central structure with $N_{\rm{H_2}} > 3 \times 10^{22}$\,cm$^{-2}$ \citep[corresponding to $A_V\sim 30$;][]{1978ApJ...224..132B,2017A&A...605L...5K}. We also include in the definition two branches, Branch-1 and Branch-8, since they are aligned  along  the main axis of the central region. 
All  other branches are considered part of the filamentary network.

Based on the column density skeleton, we estimate the length of G351 to be $\sim$4.6\,pc, assuming an inclination of $0 \deg$ against the plane of the sky. This estimate is also confirmed by the \textit{Spitzer} 8\,$\mu$m data. Clump-1, the most evolved in G351, is found roughly in the middle of the filament ($\sim$2.5\,pc starting from the northern end) and  divides G351 into two parts: the northern part, which is cold and dense, and is clearly detected in emission at FIR wavelengths and absorption in the mid-IR, and the southern part, which is clearly detected in the dust emission in the FIR and hosts more evolved sources. 
\begin{table*}
\centering
\caption{Width of the G351 filament measured from the average profile in each region and deconvolved for the  beam of their corresponding dataset.}\label{tab3}
\begin{tabular}{ccccccc}
\hline
Region&\multicolumn{2}{c}{SABOCA}&\multicolumn{2}{c}{SPIRE-250\,$\mu$m}&\multicolumn{2}{c}{PACS-160\,$\mu$m}\\
&pc&flag\tablefootmark{a}&pc&flag\tablefootmark{a}&pc&flag\tablefootmark{a}\\
\hline

a&$0.06\pm0.02$&2&   $0.10          \pm0.03$  &1& $0.11       \pm           0.02$&1\\ 
&$0.10\pm0.02$&1&    $0.25 \pm           0.05$&1& $ 0.23            \pm           0.05$&1 \\ 
&$0.14\pm0.03$&1&    $0.20           \pm0.03$ &1& $0.21            \pm 0.03$&1\\
&$0.3\pm0.2$&2&      $0.3           \pm0.1$   &2& $0.3            \pm           0.1$&2\\
&$0.19\pm0.02$&1&    $0.23           \pm0.03$ &1& $0.20            \pm           0.02$&1\\ 
b&$0.113\pm0.007$&1& $0.14           \pm0.01$ &1& $0.12            \pm           0.01$&1\\ 
&$0.13\pm0.01$&1&    $0.17           \pm0.01$ &1& $0.18            \pm           0.02$&1\\ 
&$0.10\pm0.01$&1&    $0.18          \pm0.01$  &1& $0.21\pm           0.02$&1 \\ 
c&$0.11\pm0.01$&1&   $0.177\pm0.007$          &1& $0.20\pm0.01$&1\\ 
&$0.12\pm0.01$&1&    $0.165          \pm0.006$&1& $ 0.17            \pm           0.01$&1 \\ 
&$0.09\pm0.02$&2&    $0.25          \pm0.04$  &1& $0.29            \pm          0.05$&1\\ 
&$0.07\pm0.04$&2&    $0.21           \pm0.02$ &1&  $0.3           \pm           0.1$&2  \\

\hline
\end{tabular}
\tablefoot{The  width is obtained with a Gaussian fit to the average profile. For the {\it Herschel} data a non-zero baseline is used.\\ 
  \tablefoottext{a}{We assign a flag of 2 to the regions along G351 for which the relative error in the width estimate is larger than 20\%.
    The reported widths are based on the fit results and deconvolved for the beam size, while the errors are that inferred from the fit.}}
\end{table*}

\begin{figure*}[!htb]
\centering
\includegraphics[width=0.9\textwidth]{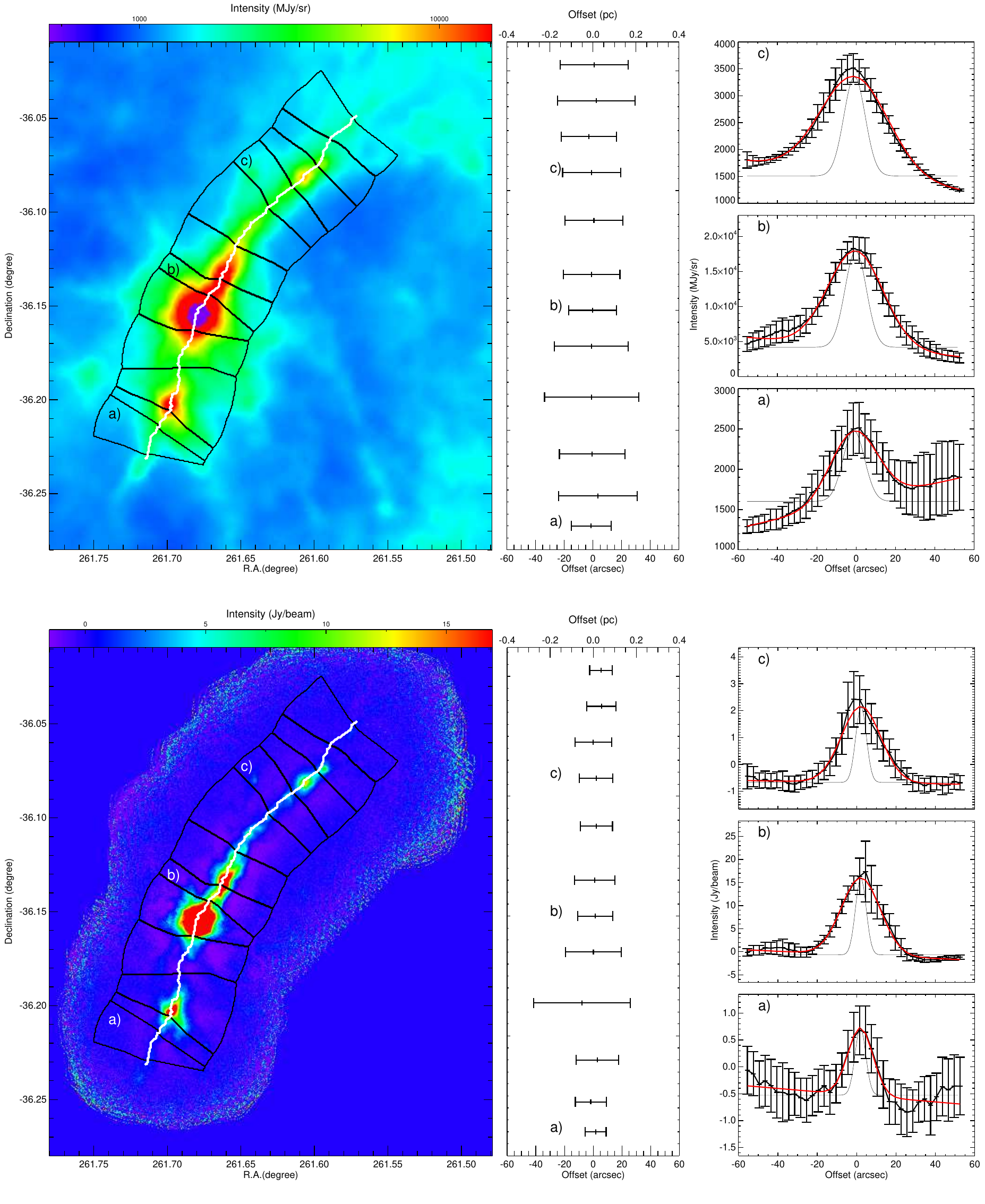}
\caption{{\bf Top:} Hi-GAL-250\,$\mu$m map of G351 (left panel). The white solid  line indicates the skeleton of the filament obtained from the Hi-GAL data. The map is divided in different regions for the estimate of the filament width along the source. In the middle panel, we show the full width at half maximum in arcsec in different regions of the filament.  The cross-spine average radial profile of the observed intensity in the  Hi-GAL-250\,$\mu$m data (black thick line, right panel) is given for three regions (a, b, and c; these are also shown in the other two panels) along G351. The solid red line is the Gaussian fit; the thin black line shows the resolution of the data. The error bars show the dispersion of the measurements in the relative region. {\bf Bottom:} As for the top panel but for the SABOCA data.
}
\label{saboca}
\end{figure*}

To estimate the width of G351 and determine how it changes along  its length, we used the photomotric maps at 350\,$\mu$m from SABOCA, 
  which have the highest angular resolution among our submillimetre data but filter out the large-scale emission, and   at 160\,$\mu$m and 250\,$\mu$m from Hi-GAL, which are sensitive also to the diffuse emission. We selected these two Hi-GAL bands because they have the closest  angular resolution (see Table\,\ref{tab2})  and are close in wavelengths to the SABOCA data.
We made the simple hypothesis that the structure is isothermal when
estimating the physical width from the photometric maps. Later we
discuss how our results change when dropping this assumption.  
We divided the maps in several 
regions (see Fig.\,\ref{saboca} for the analysis on the SABOCA data and on the 250\,$\mu$m map) and 
studied the radial profiles as a function of the distance from the dust skeleton along the length of G351, in particular to investigate variations of its width. The regions have similar lengths of $\sim0.3-0.4$\,pc along the spine 
 and they are selected to cover the entire length of the main filamentary structure.
Results are shown in Fig.\,\ref{saboca} and Fig.\,\ref{saboca_spire} and in Table\,\ref{tab3}. The profile of G351 is spatially resolved in all  wavelengths adopted for this analysis. The profiles obtained from Hi-GAL are consistently broader than those derived from the SABOCA data, likely because of filtering effects. In the SABOCA data, the width, $w$, of the filament is well resolved everywhere along the skeleton (see right panel of Fig.\,\ref{saboca}) with a median value of $\sim 0.11$\,pc and a standard deviation of 0.07\,pc. The Hi-GAL profiles are typically 0.2\,pc broad ($w_{160}=0.19\rm{\,pc}\pm 0.05\rm{\,pc}$, $w_{250}=0.21\rm{\,pc}\pm 0.06\rm{\,pc}$).
The radial profiles for all analysed regions are shown in  Fig.\,\ref{saboca_spire} at three wavelengths (350\,$\mu$m from SABOCA, 250\,$\mu$m and   160\,$\mu$m from Hi-GAL).
Interestingly,  the width of the filament is essentially constant along its 4.6\,pc length for all of the data and all regions, and we do not see any significant deviation from Gaussian shape. Moreover, the widths derived using different Hi-GAL bands are consistent with each other. We also  derived the width of the filament using the C$^{18}$O\,(2--1) emission integrated in the velocity range $-3.5\pm0.5$\,km\,s$^{-1}$ (see Sect.\,\ref{sec_co}) and found that the width of the diffuse medium associated with G351 is slightly broader ($w_{\rm{C^{18}O}}=0.37\pm0.06$\,pc) but also constant along its full extent.

To investigate the impact on our analysis of the assumption that the dust
  emission is isothermal, we combined our temperature map with the  250\,$\mu$m thermal continuum emission and  derived a column density distribution at 18\arcsec resolution directly from the 250\,$\mu$m data \citep[see for example][]{2016A&A...592A..54A}. We then estimated the width of the filament in the 12 regions described above using the same methodology applied to the photometric maps. Results from both methods are in agreement within the reported errors (Fig.\,\ref{temp}), demonstrating that our assumption of isothermal emission does not influence the results even though the temperature along the filament is not constant. To summarise, the main filamentary body G351 is a structure with an average column density of $8\times 10^{22}$\,cm$^{-2}$,  an aspect ratio of $\frac{4.6}{0.2}\sim 23$, and an almost constant width of $\sim0.2$\,pc.

\begin{figure}[!htb]
\includegraphics[width=0.9\columnwidth]{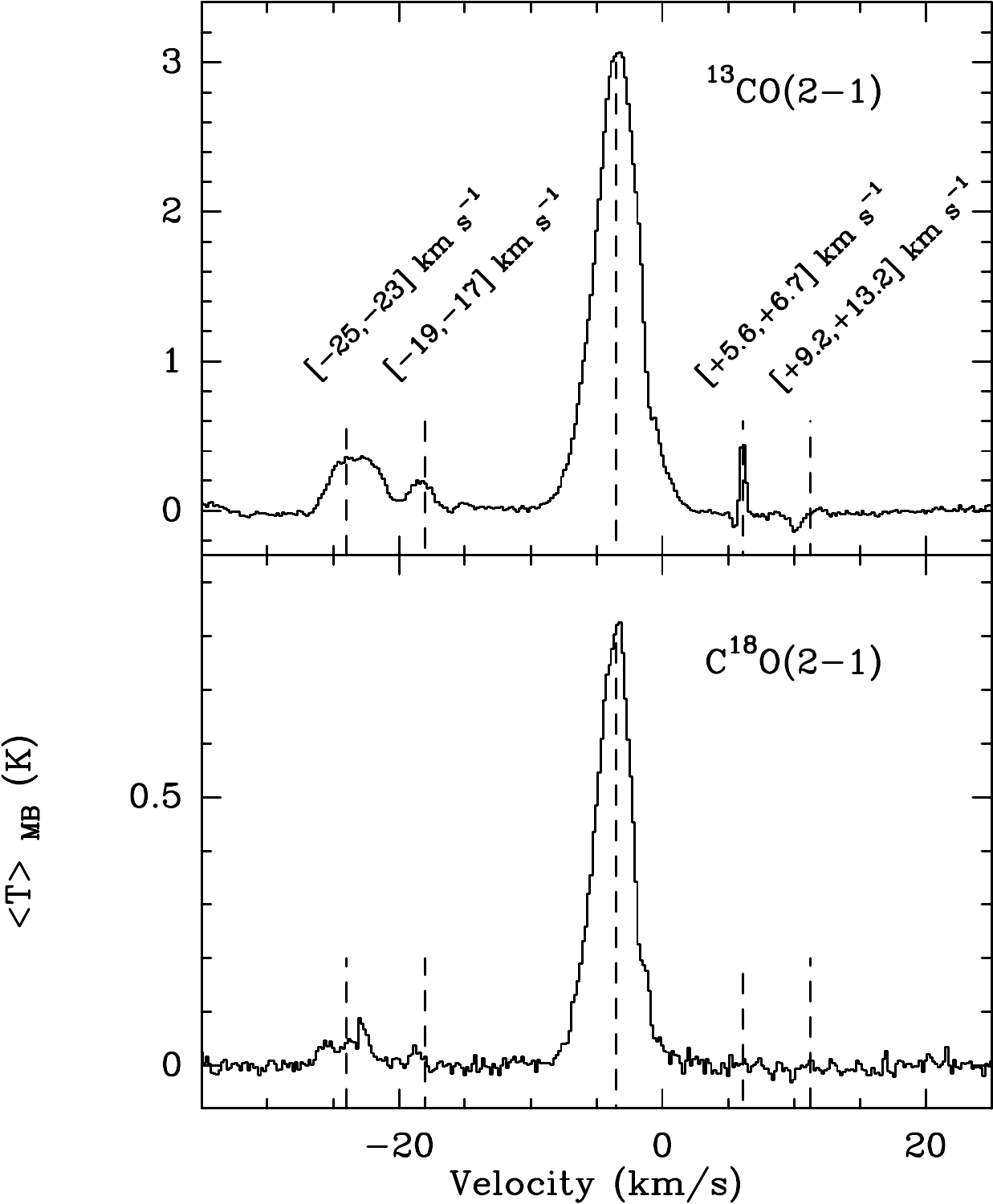}
\caption{Averaged $^{13}$CO\,(2--1) (top) and C$^{18}$O\,(2--1) (bottom) spectra over the full extent of the APEX map. The dashed lines indicate $\varv_{\rm{LSR}}$ of $-18$\,km\,s$^{-1}$, $-24$\,km\,s$^{-1}$, $-3.5$\,km\,s$^{-1}$,  $+6$\,km\,s$^{-1}$,  $+11$\,km\,s$^{-1}$. 
  The velocity ranges used to obtain the integrated intensity maps shown in Fig.\,\ref{clouds} are also shown.}
\label{spectra}
\end{figure}

\subsection{CO  isotopologue  emission}\label{sec_co}

Figure\,\ref{spectra} shows the  $^{13}$CO\,(2--1) and C$^{18}$O\,(2--1) spectra averaged over the full region mapped with APEX.
Figure\,\ref{clouds} shows the distribution of the   $^{13}$CO emission separately integrated over the velocity intervals of  the five different
velocity components clearly visible in the integrated $^{13}$CO spectrum
(peaked around $-24$\,km\,s$^{-1}$, $-18$\,km\,s$^{-1}$, $-3.5$\,km\,s$^{-1}$, $+6$\,km\,s$^{-1}$, and $+11$\,km\,s$^{-1}$). The weak  component at $\sim+11$\,km\,s$^{-1}$ is seen mostly in the channel maps since the emission is strongly confined to a small region in the north-west of G351 (see Fig.\,\ref{clouds}e).  Clearly G351 and the majority of the dark patches seen in extinction in the mid-IR and in emission in the Hi-GAL data belong to the same molecular complex with a typical velocity of $\sim -3$\,km\,s$^{-1}$ (see also Fig.\,\ref{fig1}). Branch-3, -4, -5, -6, and -8 are detected in both $^{13}$CO and C$^{18}$O, while 
 Branch-2  can be seen only in the $^{13}$CO integrated emission map (see Fig.\,\ref{clouds}).
The other components are  not associated with the  molecular environment of G351.
The emission at $\sim-24$\,km\,s$^{-1}$ is associated with Midcourse Space Experiment (MSX) IRDC G351.64--00.46 \citep{2006ApJ...639..227S} on the south-west of the main filament (see Fig.\,\ref{clouds}). Likely, the 8\,$\mu$m bright emission on the south of MSX IRDC G351.64--00.46 is associated with the same molecular environment  since in Paper\,I we found CO isotopologue emission centred at $\sim -22$\,km\,s$^{-1}$ at their positions 9, 10, and 12 (see their Figure\,3).

In the following sections we derive the mass of G351 and of the full
network of branches using the dust maps and the C$^{18}$O\,(2--1) data
(Sect.\,\ref{sec_mass}), and analyse the velocity field in the
structure (Sect.\,\ref{sec_vel}) and its dynamical state
(Sect.\,\ref{sec_virial}).  We focus only on the optically thin
C$^{18}$O\,(2--1) data, avoiding high opacity effects probably
affecting the $^{13}$CO spectra. Indeed we verified that C$^{18}$O has
low optical depths on the dust condensations in G351 comparing with
C$^{17}$O\,(2--1) data from Paper\,I. The observed ratio of the
integrated intensities of C$^{18}$O to C$^{17}$O ranges between 2.4
and 5 under the assumption of opthically thin  C$^{17}$O emission, implying moderate opacities for C$^{18}$O of 0.2--0.5 even
towards the most massive clumps in G351 and suggesting lower optical
depths in the rest of the region.

\subsection{Mass distribution of the filamentary system}\label{sec_mass}

We computed the mass of the filament and the network from the Hi-GAL data and from the C$^{18}$O map assuming thermal equilibrium between dust and gas and using the dust temperature map from {\it Herschel} as input to estimate the C$^{18}$O column density. 
Around the saturated region we measure a maximum temperature of $\sim$30\,K. We adopted this value as a lower limit for the effective temperature on Clump-1  given the presence of an H{\sc ii} region and a hot core. This assumption translates into an lower limit to the real mass for the estimate based on C$^{18}$O. 
Assuming a CO abundance of 
$10^{-4}$ \citep{2017A&A...603A..33G}, a $^{16}$O/$^{18}$O ratio of $58.8 D_{\rm{gc}}+37.1$, \citep{1994ARA&A..32..191W}, and a Galactocentric distance, $D_{\rm{gc}}$,  equal to 7.4\,kpc \citep[][revised for $d=1$\,kpc]{2014A&A...570A..65G},  the total mass of the system including the network of filaments derived from C$^{18}$O is $\sim 2100$\,M$_{\odot}$. 
For the main body ($N_{\rm {H_{2}}}\geq 3\times 10^{22}$\,cm$^{-2}$), 
we estimated a mass of 
$\sim 1200 \rm{M}_{\odot}$.  
We do however expect  higher temperatures close to protostellar sources in particular in the proximity of Clump-1 \citep[see][]{2011A&A...530A..12L}. 
Adopting $T=50\,K$ for the saturated region \citep[see][for typical temperatures of similar sources at comparable resolutions]{2017A&A...603A..33G,2018MNRAS.473.1059U}, the mass estimates based on C$^{18}$O becomes $\sim2200$\,M$_\odot$ for the whole
region and $\sim1300$\,M$_\odot$ for the main filament.

We also derived the mass of G351 ($\sim\,$1870\,M$_{\odot}$) and that of the whole region including the network ($\sim\,$2300\,M$_{\odot}$) using the 
Hi-GAL column density map. These  estimates do not include the saturated  region close to Clump-1. Therefore we also attempted to estimate the mass hosted in the saturated region using the ATLASGAL data assuming a temperature of 30\,K as for C$^{18}$O and the same dust opacity law used for the Hi-GAL data. This assumption leads to a mass of $\sim 190$\,M$_\odot$. We note that the values  reported in Paper\,I for the same clump were given for temperatures of 10\,K, 25\,K,  and 35\,K and for a different opacity law. 
We finally note  that  the mass estimate for the network derived from the dust is likely a lower limit  due to the difficulty to disentangle the low-contrast branches from the background in a quite extended region of the cloud. Without any background subtraction 
the total mass of the network including G351  increases to $\sim 4560$\,M$_\odot$.

Although the  masses based on the dust and on C$^{18}$O are in 
  good agreement, the  C$^{18}$O estimate for the main structure is  lower than that derived from the dust. This is could be due to opacity effects and self-absorption in the southern region of G351, where the most massive clumps reside, and to depletion in the youngest and coldest northern part
  \citep[see][for depletion in IRDCs, and Sabatini et al. (in prep.) for a study of depletion in G351] {2011ApJ...738...11H}.  Also deviation from the local thermal equilibrium assumption can play a role in this discrepancy.

 To summarise, we estimate that the total mass of the network including G351 is between 2600\,M$_\odot$ and 4600\,M$_\odot$ including the mass of the saturated region; there is a a large uncertainty due to the difficulty to disentangle the low-contrast branches from the background. The main filamentary body has a total mass of 1600--2100\,M$_\odot$.
 Our analysis confirms that the network of branches  surrounding G351 harbours a large reservoir   of  gas  and dust which could still be accreted onto the clumps detected along the main filament. This reservoir accounts for about 20\,\% of the total mass if we adopt the estimate of 2600\,M$_\odot$ for the network based on the dust and subtracting a background, but increases to up to $\sim 60\%$ if no background subtraction is done. Using the C$^{18}$O data, the mass of the network is $\sim40\%$ of that of the main substructure.

\begin{figure}
\centering
\centering\includegraphics[width=0.85\columnwidth,angle=-90]{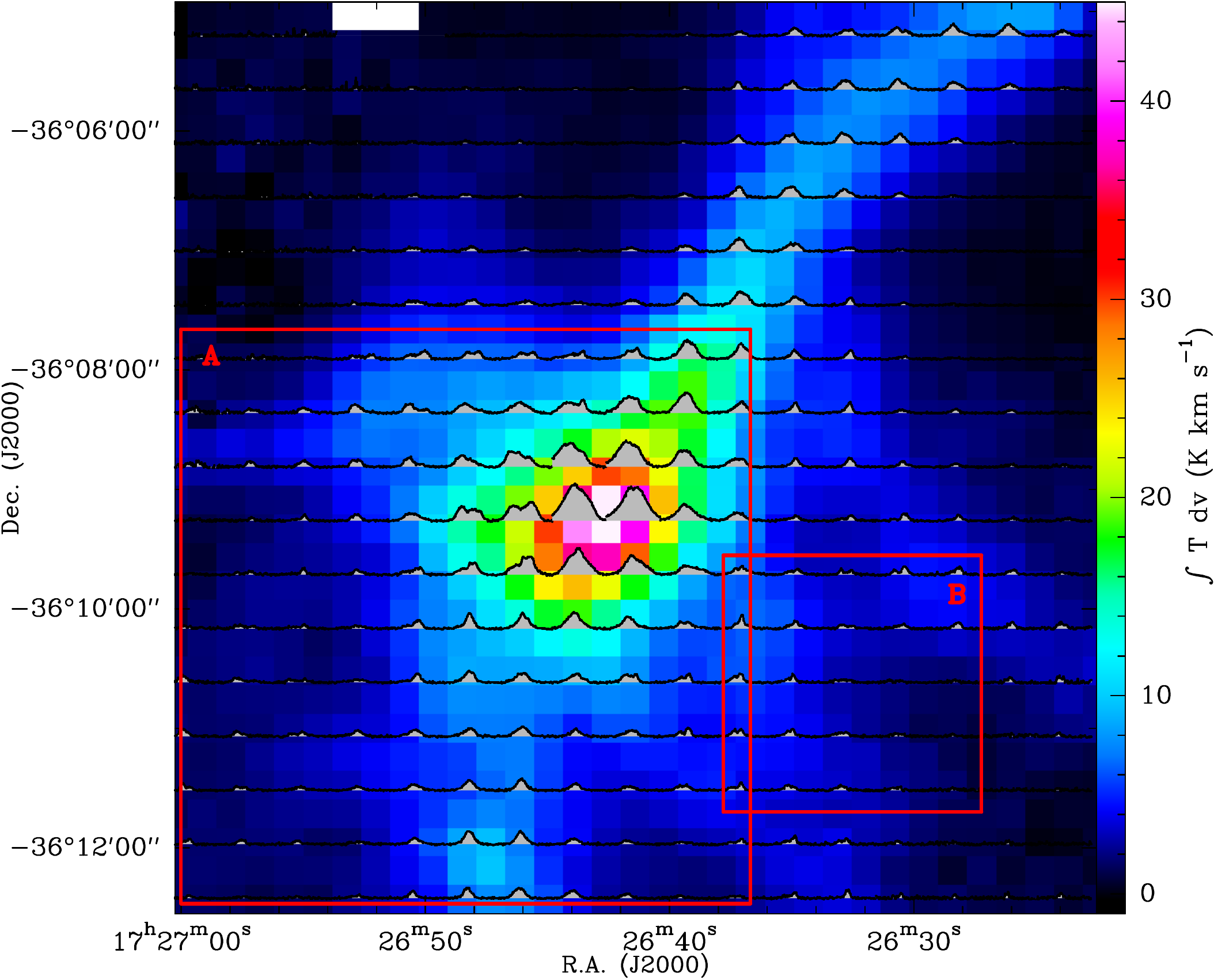}
\caption{Map of the C$^{18}$O\,(2--1) transition overlaid on the integrated emission of the same line  in the velocity range  $[-6,-1]$\,km\,s$^{-1}$. The red boxes indicate the regions shown in Fig.\,\ref{c18o_zoom}.}
\label{c18o_spectra}
\end{figure}

\begin{figure}
\centering
\subfigure[][]{\includegraphics[width=0.45\textwidth,angle=-90]{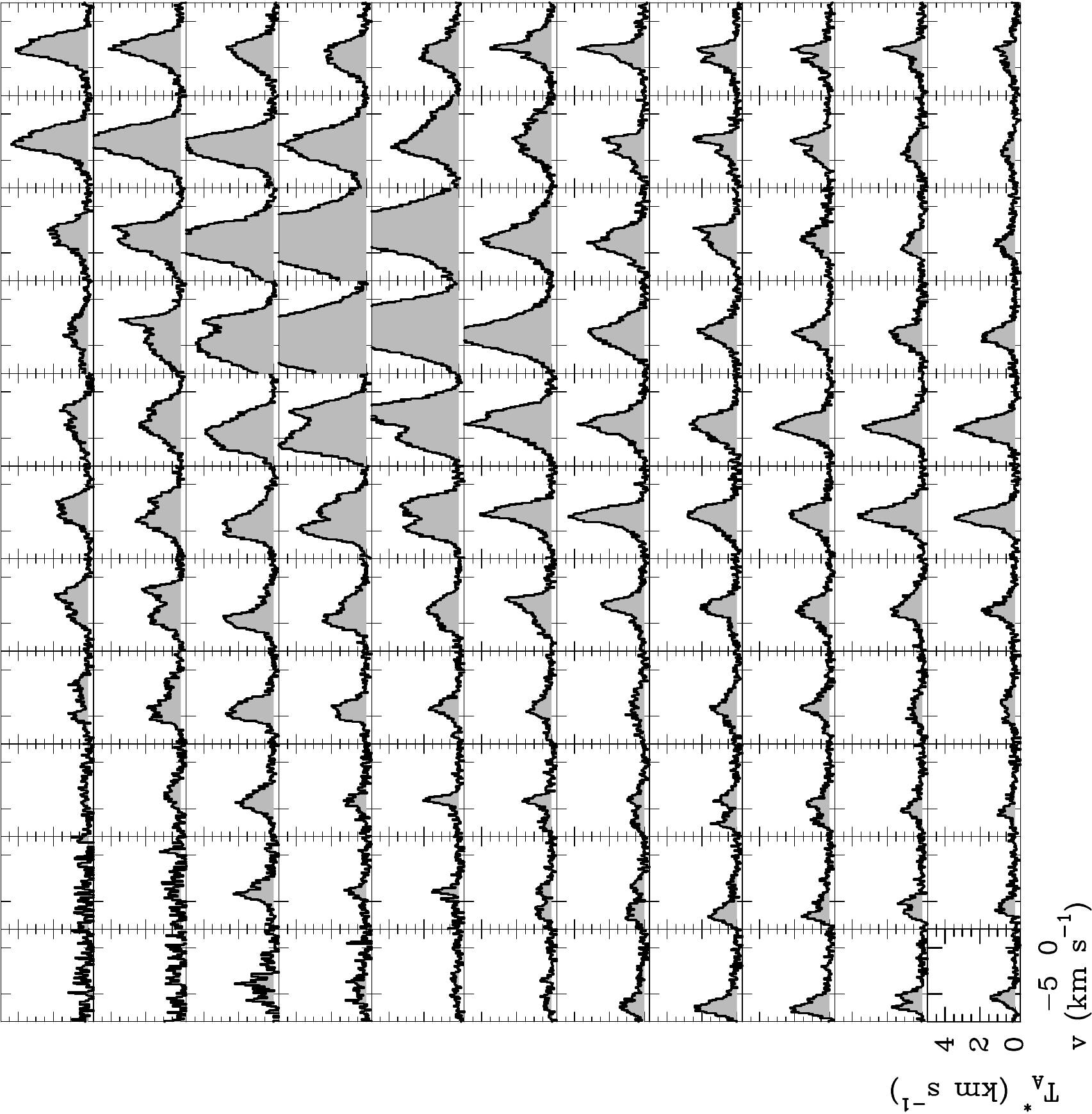}}
\subfigure[][]{\includegraphics[width=0.45\textwidth,angle=-90]{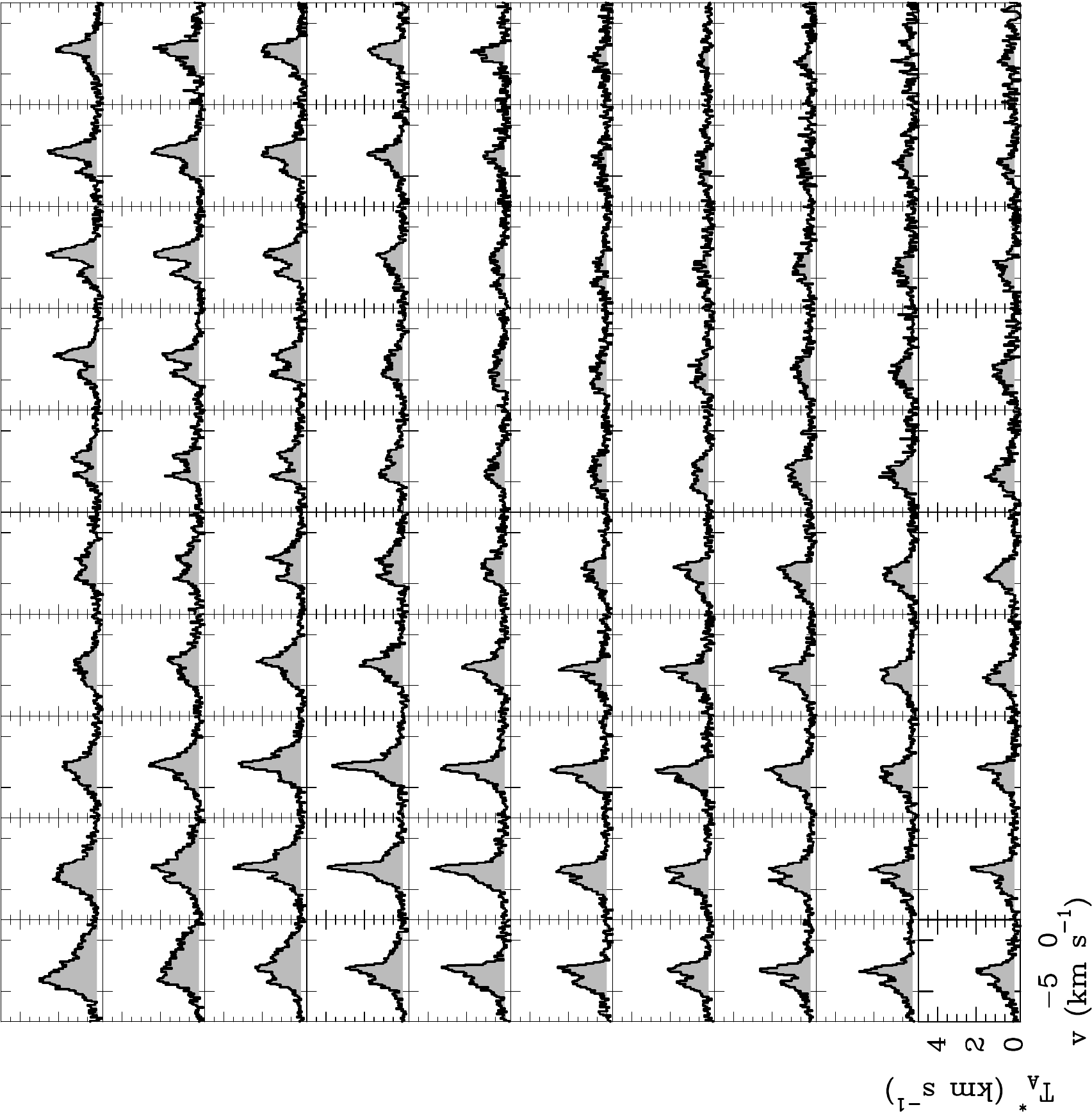}}
\caption{C$^{18}$O\,(2--1) spectral map in the two regions denoted with A (panel a) and B (panel b) on Fig.\,\ref{c18o_spectra}. A double-peaked profile is detected at several positions in the low-density gas.}
\label{c18o_zoom}
\end{figure}

\section{Velocity structure}\label{sec_vel}

In this section, we explore the gas kinematics along and across the
filament G351 and in the network of branches using the centroid velocity ($V_{\rm{lsr}}$) derived from the
C$^{18}$O emission. 
In Fig.\,\ref{chanmap} we show the C$^{18}$O channel map emission associated with the region.   One can  notice
that the emission in the  southern region is confined within the velocity range $[-4.5,- 3.4]$\,km\,s$^{-1}$,
while the northern part emits  between $-3.9$\,km\,s$^{-1}$ and $-2$\,km\,s$^{-1}$. This is due to a gradual shift of the C$^{18}$O emission for the lines of sight  from the southern to the northern portion of G351 as can be seen in Fig.\,\ref{moments}, where we show the C$^{18}$O first moment map  (see also discussion in Sects.\,\ref{sec_dv_co} and \ref{sec_disc}). 
This difference in velocity between the northern and southern part was  also reported in Paper\,I based on  N$_2$H$^+$ data.
We note that along the spine of G351, 
C$^{18}$O is  reasonably fit with a single line component except  towards IRAS\,17233--3606, where optical depth and outflow contamination is evident (Fig.\,\ref{c18o_spectra}). On the other hand, two velocity components are clearly detected in the low-density gas in a large portion of the mapped region (Fig.\,\ref{c18o_zoom}).  These two components are also clearly detected  in the northern part of G351  in the region from  Clump-6 to Clump-7 (see Fig.\,\ref{spectra_perp}) where a velocity gradient perpendicular to the main axis  was detected in the N$_2$H$^+$\,(1--0) data in Paper\,I (see their Fig.\,8). A careful analysis of the isolated hyperfine of the N$_2$H$^+$\,(1--0) line shows that a double-peaked profile is also seen at some positions along the spine of G351 (see Fig.\,\ref{n2hp_spectra}), and we speculate that the previously reported velocity gradient is  due to these double velocity components. However, the isolated N$_2$H$^+$ hyperfine
is detected only with low-signal-to-noise ratio and we refrain from a further analysis of these data.

\subsection{Large-scale velocity structure}\label{sec_dv_co}

\begin{figure*}
  \includegraphics[width=0.9\textwidth]{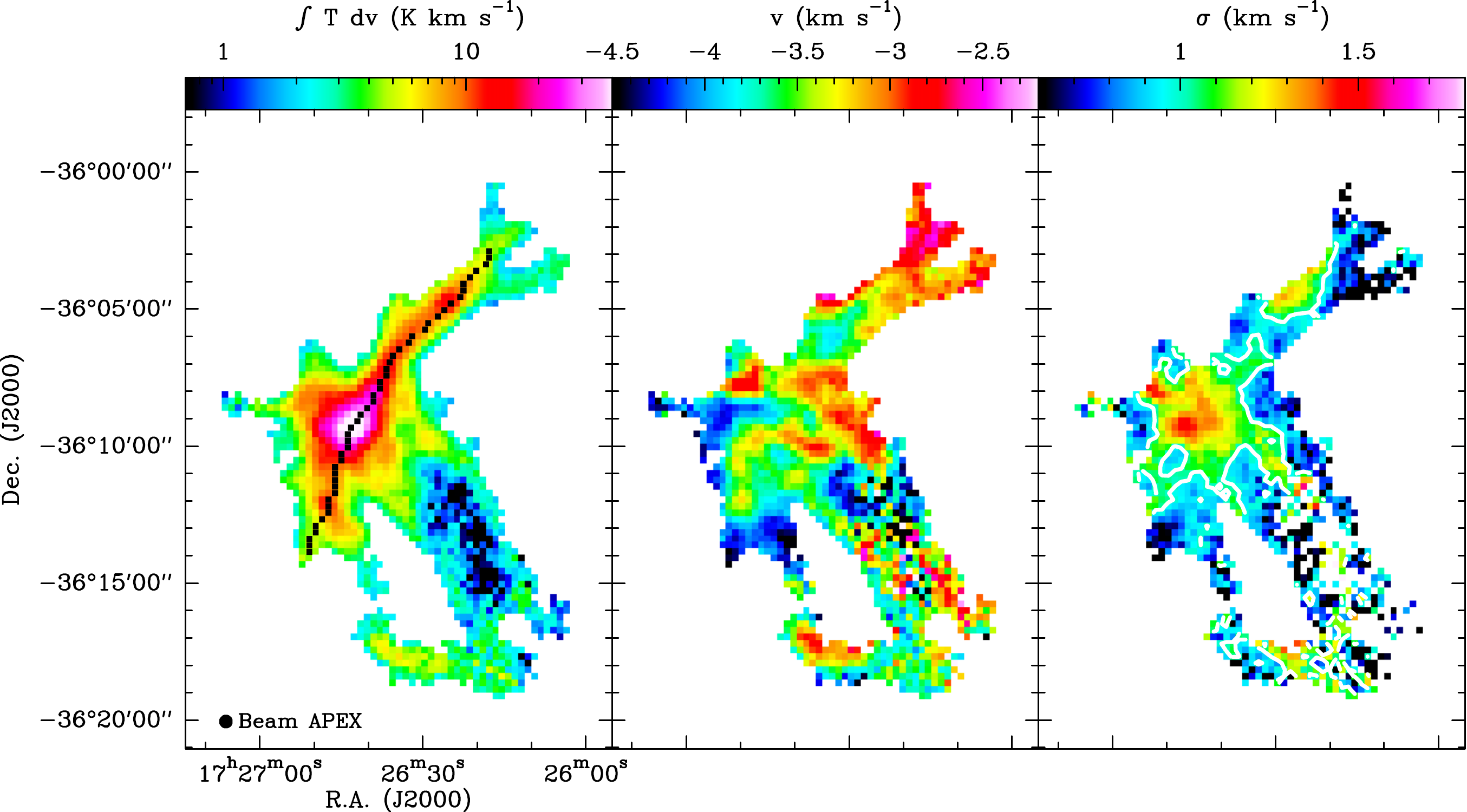}
  \caption{Zeroth (integrated intensity, left panel), first (velocity, middle panel), and second (velocity dispersion, right panel) velocity moments of C$^{18}$O\,(2--1) in the velocity range  $[-6,-1]$\,km\,s$^{-1}$. The black squares in the left panel indicate the spine of G351. The white contour in the right panel represents a velocity dispersion of 1\,km\,s$^{-1}$.}
      \label{moments}
\end{figure*}

In Fig.\,\ref{moments} we show the distribution of the first
  three  moments obtained from the C$^{18}$O\,(2--1) line in
  the velocity range $[-6,-1]$\,km\,s$^{-1}$. The overall velocity distribution of the
  C$^{18}$O\,(2--1) transition is in good agreement with that of N$_2$H$^+$ presented in Paper\,I.
  We confirm the velocity gradient detected in the northern part of G351 and also the
  gradients detected towards Clump-1 and associated with  multiple molecular outflows \citep[e.g.][]{2009A&A...507.1443L}.
  Typically, C$^{18}$O traces lower velocities than N$_2$H$^+$, but this is expected if
  N$_2$H$^+$ is associated with the inner part of the filament and the clumps.
  We do not detect any
  clear velocity gradient from the branches towards the dust clumps in
  the filament although the emission in several branches is
  systematically red-shifted with respect to the clumps in the main
  filament (Fig.\,\ref{moments}, mid panel).  The second moment map
  (Fig.\,\ref{moments}, right panel) shows that the velocity
  dispersion, $\sigma$, is relatively constant along G351 on scales of
  $\sim 30\arcsec$ ($\sim 0.2$\,pc) with a typical value of $\sim
  1$\,km\,s$^{-1}$ against the thermal velocity dispersion of
  $\sim$0.2--0.3\,km\,s$^{-1}$ at a typical (at least in the mid-IR
  dark part of G351; see Fig.\,\ref{fig4}) temperature of
  10\,K--20\,K.  This velocity dispersion is also confirmed by the
  N$_2$H$^+$\,(1--0) data analysed in Paper\,I (see their Fig.\,8).
  However, the multiple velocity components detected in
  several C$^{18}$O spectra (see Sect.\,\ref{sec_vel}) could influence
  the estimate of the velocity moments.

\begin{figure*}
\centering
\includegraphics[angle=-90, width=0.97\textwidth]{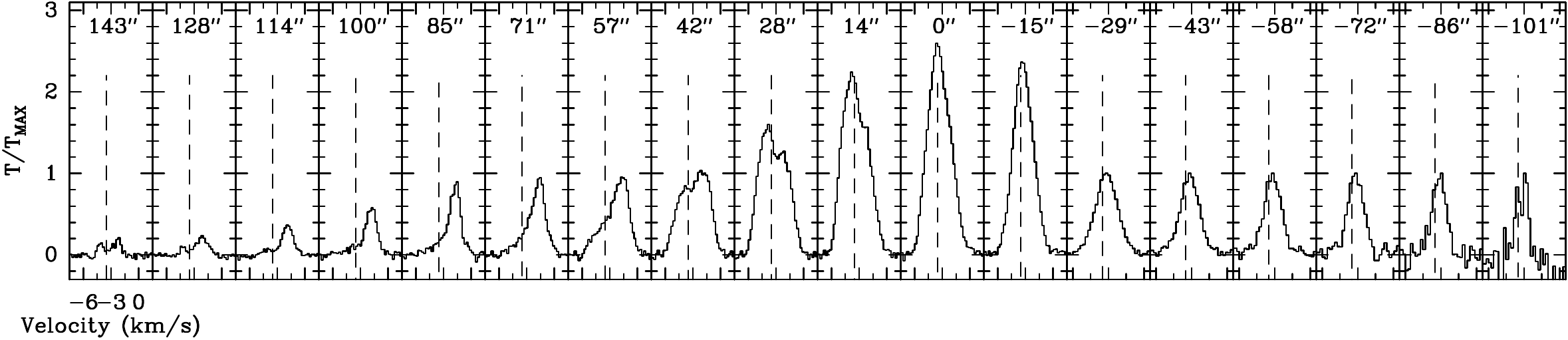}
\caption{C$^{18}$O spectra across the main axis of G351   in the most quiescent northern part of the filament. Spectra are collapsed parallel to the main axis. Each spectrum is  normalised to its peak intensity value for visualisation. The numbers in each panel indicate the distance from the spine of G351. The label 0\arcsec\, denotes the spine. The dashed line indicates a reference velocity of $-3.5$\,km\,s$^{-1}$ (see Fig.\,\ref{spectra}).}
\label{spectra_perp}
\end{figure*}

To probe the large-scale velocity distribution, we extract
C$^{18}$O\,(2--1) spectra along the main axis of the filament, in the
perpendicular direction, and along Branch-6 (one of the branches in
the network; see Fig.\,\ref{fig1}) using the skeletons obtained from
the 250\,$\mu$m data and from the H$_2$ column density map after
reprojecting them on the C$^{18}$O data. We then extracted spectra at
each position along the three structures (see
Figs.\,\ref{spectra_perp}, and \ref{br7}). The individual spectra are
analysed in CLASS using the GAUSS fit routine.  As noted in the
previous section, the C$^{18}$O data along the spine are well fit with
a single velocity component in the majority of the positions. The
median dispersion velocity along this direction is $\sim
1.1\pm0.5$\,km\,s$^{-1}$, and this value is not biased by broad
profiles associated with active clumps, since it decreases to $\sim
1.0\pm0.1$\,km\,s$^{-1}$ if we exclude the region around Clump-1 and
-2.  In Fig.\,\ref{c18o_dust} we present the velocity field (plotted
as velocity centroid against linear distance in arcsec along the
spine). In the top panel of Fig.\,\ref{c18o_dust}, we compare the
velocity centroid to the {\it Herschel} 500\,$\mu$m dust brightness,
the closest in angular resolution to the APEX observations among the
Hi-GAL data. On top of an overall velocity gradient, C$^{18}$O
exhibits a remarkable oscillatory pattern. This pattern is
discussed in more detail in Sect.\,\ref{sec_disc}.

We also investigated the C$^{18}$O spectra across the width of the
main filament. We first selected the most quiescent northern part of
G351 and collapsed the data along the direction parallel to the
skeleton to increase the signal-to-noise ratio at each position across
the spine (Fig.\,\ref{spectra_perp}).
It can be clearly seen that the
C$^{18}$O profiles get narrower at increasing distance from the
spine. Moreover, at the outermost positions on the east side of G351 a
double-peak profile is detected in the C$^{18}$O\,(2--1) line that
cannot be resolved in the denser part of G351. We stress that
this analysis is performed in the region in which a velocity gradient across the width of G351 is seen in
in C$^{18}$O and  N$_2$H$^+$ (see Fig.\,\ref{moments} and Paper\,I).
It is unlikely that optical depth effects may alter the C$^{18}$O line profile at the outer positions, since the double peaks
are detected in relatively low column density gas (see Fig.\,\ref{c18o_spectra}). Moreover, low opacities are inferred towards the dust clumps
by comparison with the C$^{17}$O spectra of Paper\, I (see Sect.\,\ref{sec_co}).
The spectra shown in Fig.\,\ref{spectra_perp} are highly reminiscent of those detected by \citet{2006A&A...445..979P} in the intermediate-mass
NGC\,2264-C clump and interpreted as the results of colliding flows.

Regarding Branch-6 (Fig.\,\ref{br7}), a second velocity component is detected at some positions along its skeleton. 
The fit results are therefore grouped together using the DBSCAN clustering algorithm\footnote{\url{http://scikit-learn.org/stable/modules/generated/sklearn.cluster.DBSCAN.html}}. A crucial input parameter for the algorithm is the number of samples in a neighbourhood for a point to be considered as a core point. We adopt a value of 2 for this parameter, which is equivalent to adopting a "friends of friends" approach  such that components that closely follow each other both in position and velocity are grouped together. The clustering results are also sensitive to the maximum distance between two samples for them to be considered as in the same neighbourhood. After testing a range in input parameters, we adopt a parameter set that agrees best with our visual verification. This result in two clusters along the spine of Branch-6: one cluster associated with its full length and a second detected only within a distance of 200\arcsec\, from G351 (Fig.\,\ref{dv_br7}). As in the case of the spine of G351, 
the first component, associated with the whole branch, 
shows an oscillatory pattern
on top of an overall velocity gradient of about 1.5\,km\,s$^{-1}$ over 600\arcsec.

\subsection{Velocity structure along the spines}\label{sec_disc}

\begin{figure}
  \centering
  \includegraphics[width=0.99\columnwidth, angle=-90]{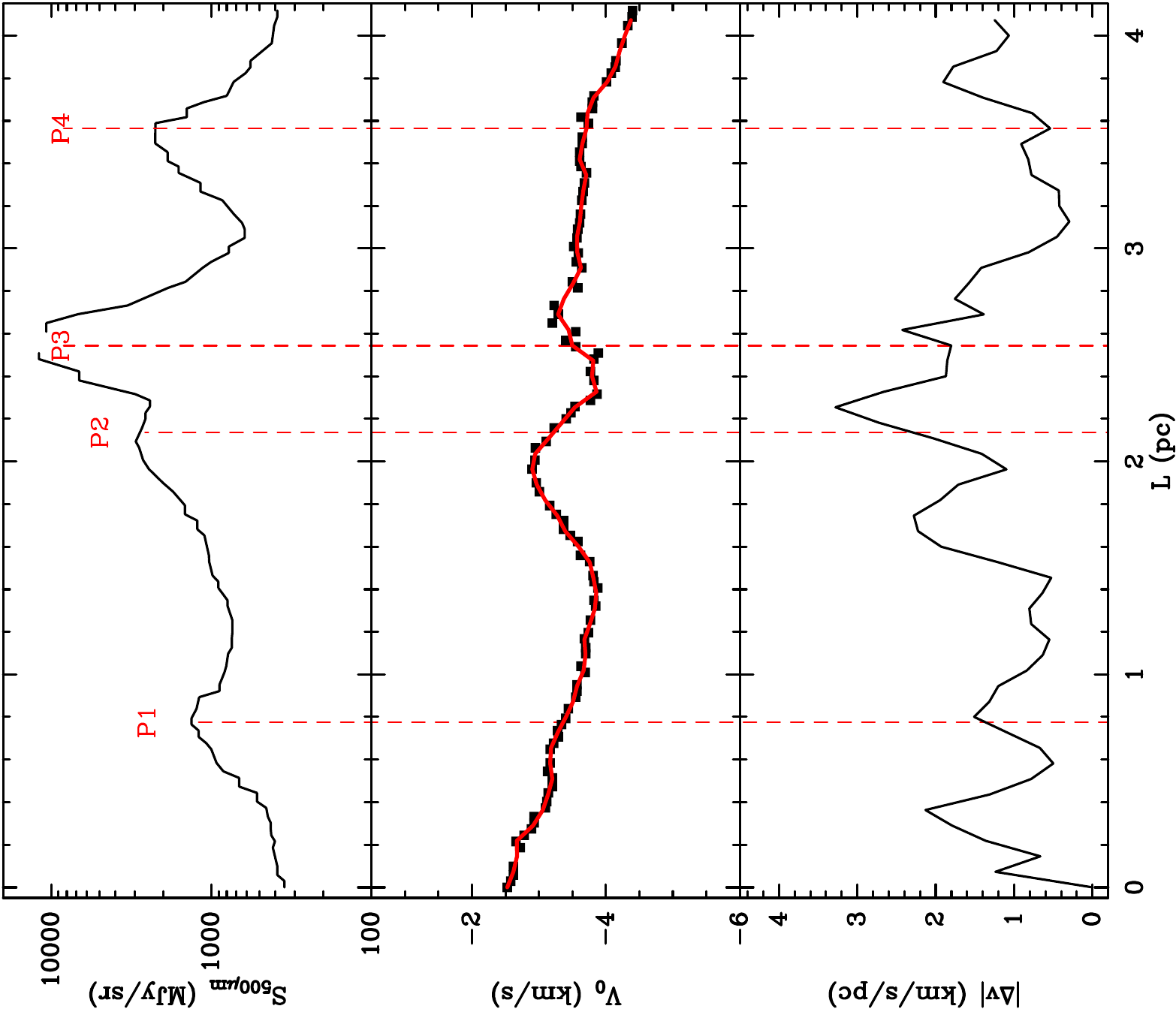}
  \caption{Top panel: Distribution of dust brightness from the Hi-GAL-500\,$\mu$m map (with 36\arcsec resolution,
    comparable with that of the line  data) as a function of position along the main spine of G351. The starting point for the skeleton is the position
    $\alpha({\rm J2000})$ = 17$^h$26$^m$17\farcs2, $\delta({\rm J2000})$ = --36$\degr$02$\arcmin$56\farcs7.
    Labels P1--P4 denote the four peaks detected in the {\it Herschel} 500\,$\mu$m dust brightness and correspond to Clump-5, Clump-2, Clump-1, and Clump-3,
    respectively. Mid-panel: Velocity centroid for C$^{18}$O\,(2--1)  as a function of position along the skeleton of G351. 
    The black squares represent the data points, the red curve obtained by resampling the data over a bin of 15\arcsec (equivalent to 0.07\,pc). Botton panel: Absolute velocity gradient
 along the spine computed over 0.07\,pc.}\label{c18o_dust}
\end{figure}

\begin{figure}
\centering
\includegraphics[width=0.48\textwidth]{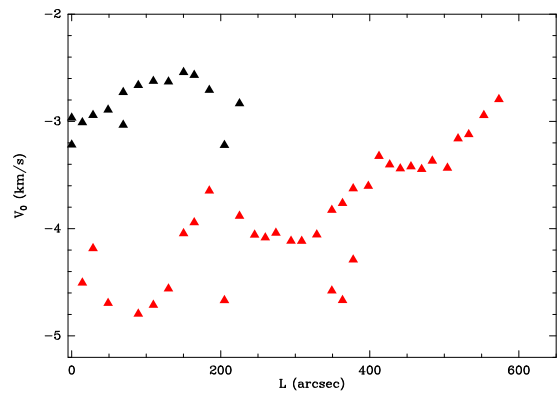}
\caption{C$^{18}$O\,(2--1) velocity centroid as a function of position along the spine of Branch-6. The red and black triangles refer to the two different velocity groups identified by the clustering algorithm.}
\label{dv_br7}
\end{figure}

Figure\,\ref{c18o_dust} shows the flux density of the 500\,$\mu$m Hi-GAL data, the 
centroid line-of-sight velocity of the C$^{18}$O\,(2--1) line, and the absolute
value of its gradient  along the spine of G351 as a function of distance along the spine.
The absolute velocity gradient is evaluated over a bin of 0.07\,pc.
As already noted in Sect.\,\ref{sec_dv_co},   velocity fluctuations are
detected along the spine of G351 and at least of Branch-6 in
C$^{18}$O (Fig.\,\ref{dv_br7}). We also note that on top of a large-scale velocity gradient of about
$\sim 1$\,km\,s$^{-1}$\,pc$^{-1}$, local velocity gradients close are developed close to dust clumps, similar
to the findings of \citet{2018A&A...613A..11W} in the massive filamentary hub SDC13.

While we note a general association between the
fluctuations and peaks of the velocity gradient and location of dense cores, we do not find a one-to-one
correlation. However, all peaks of the velocity gradient are found in within $\sim0.2$\,pc from
dust peaks. Given the typical size of the clumps in G351 ($\sim0.1-0.2$\,pc, see Table\,3 in Paper\,I), local
motions could influence the velocity field in the vicinity of the clumps.
\citet{2018A&A...613A..11W} interpreted  the correlation
between  the  pattern of the velocity field and the dust peaks as a signature of accretion  from  the  parent
filament. However, these velocity variations can have a different
interpretation than the steady accretion.  For example \citet{2016MNRAS.455.3640S}  found similar velocity shifts in their simulations and
determined that they are ascribed to transient motions.

In one of the very few cases where a good agreement
between velocity and density peaks has been found (low-mass filament
L1571, \citealt{Hacar2011}), it has been argued that this is a
consequence of filament fragmentation via accretion along
filaments. These authors modelled such velocity oscillations as sinusoidal
perturbations caused by mass inflow along filaments forming
cores. Such motions are expected to create a $\lambda/4$ shift between
core density and velocity field.  As discussed in
\citet{2014MNRAS.440.2860H}, this applies to an idealistic
case. Projection effects can obscure the actual inclination angle of
the filament, and thus lead to different velocity signatures. Recent
simulations \citep{2016MNRAS.463.4301H,2017ApJ...834..202G} have
started to model the observed velocity oscillations as a natural
consequence of filaments that start out in hydrostatic equilibrium and
undergo subsonic gravitational fragmentation. Additional effects such
as inclination and geometric bents would produce different velocity
oscillatory patterns.

Unlike the low-mass filament L1571 in which velocity oscillations have
been observed, fluctuations in G351 are highly supersonic.  The peak
to valley variations in the velocity profile are $< 2$\,km\,s$^{-1}$
in either of the two tracers. The free-fall velocity implied for the
most massive core in the filament, a 400\,M$_{\odot}$ hot core of
radius $\sim 0.1$\,pc (see Tables 3 and 4 of
\citealt{2011A&A...533A..85L}) is $\sim 6$\,km\,s$^{-1}$. The
rotational energy implied by the velocity gradient (see
Fig.~\ref{c18o_dust}) is negligible compared to the kinetic
energy. Therefore, there have to be other factors that significantly
influence the gas dynamics. For example, only a fraction of total
velocity variation would be identified for a filament inclined at an
angle with respect to the observer \citep{2014MNRAS.440.2860H}.
Moreover, strong magnetic fields can also significantly slow down
accretion (Pillai et al. in prep).

\section{Dynamical state of G351}\label{sec_virial}

  In the previous section, we reported that the velocity dispersion in
  G351 is relatively constant ($\sim 1$\,km\,s$^{-1}$, see
  Fig.\,\ref{moments} right panel, and Sect.\,\ref{sec_dv_co}) at the
  resolution of our C$^{18}$O and N$_2$H$^+$ observations. However, a
  double velocity profile is detected in the majority of positions in
  the low column density gas (Figs.\,\ref{c18o_zoom} and
  \ref{spectra_perp}) and we have indications that these double
  component is also seen along the spine of G351 in the isolated
  hyperfine of N$_2$H$^+$\,(2--1) (see Fig.\,\ref{n2hp_spectra}).
  Unfortunately, our data do not allow us to analyse these components
  in further detail. The median value of the velocity dispersion
  derived on the spine of G351 from C$^{18}$O is 1.0\,km\,s$^{-1}$
  (excluding the positions close to Clump-1 and -2).
  This value is higher than that found in nearby filaments by \citet{2013A&A...553A.119A}; however
  our estimate is close to measurements in more massive objects \citep[e.g.][]{2010A&A...520A..49S,2018arXiv180807499M}.
 Moreover, we point out that our measurement can be
  biased by  high opacity  and dilution  effects, which  decrease the  estimate of
 $N_{\rm{H_2}}$, or by the presence of multiple components.

Using a velocity dispersion of 1\,km\,s$^{-1}$, we derive a virial
mass per unit length $M_{\rm {line, vir}}=2 \sigma^2_{tot}/G\sim
470$\,M$_\odot$\,pc$^{-1}$ where $\sigma_{tot}$ is the total velocity
dispersion (thermal plus non-thermal) and $G$ the gravitational
constant.  Considering a total mass of $1600-2100$\,M$_\odot$ for the
main filament G351 and a length of 4.6\,pc (see Sects.\,\ref{sec_dust}
and \ref{sec_mass}), the mass per unit length, $M_{\rm {line}}$, is
$350-450$\,M$_\odot$\,pc$^{-1}$.

\begin{figure}
\centering
\includegraphics[width=0.8\columnwidth]{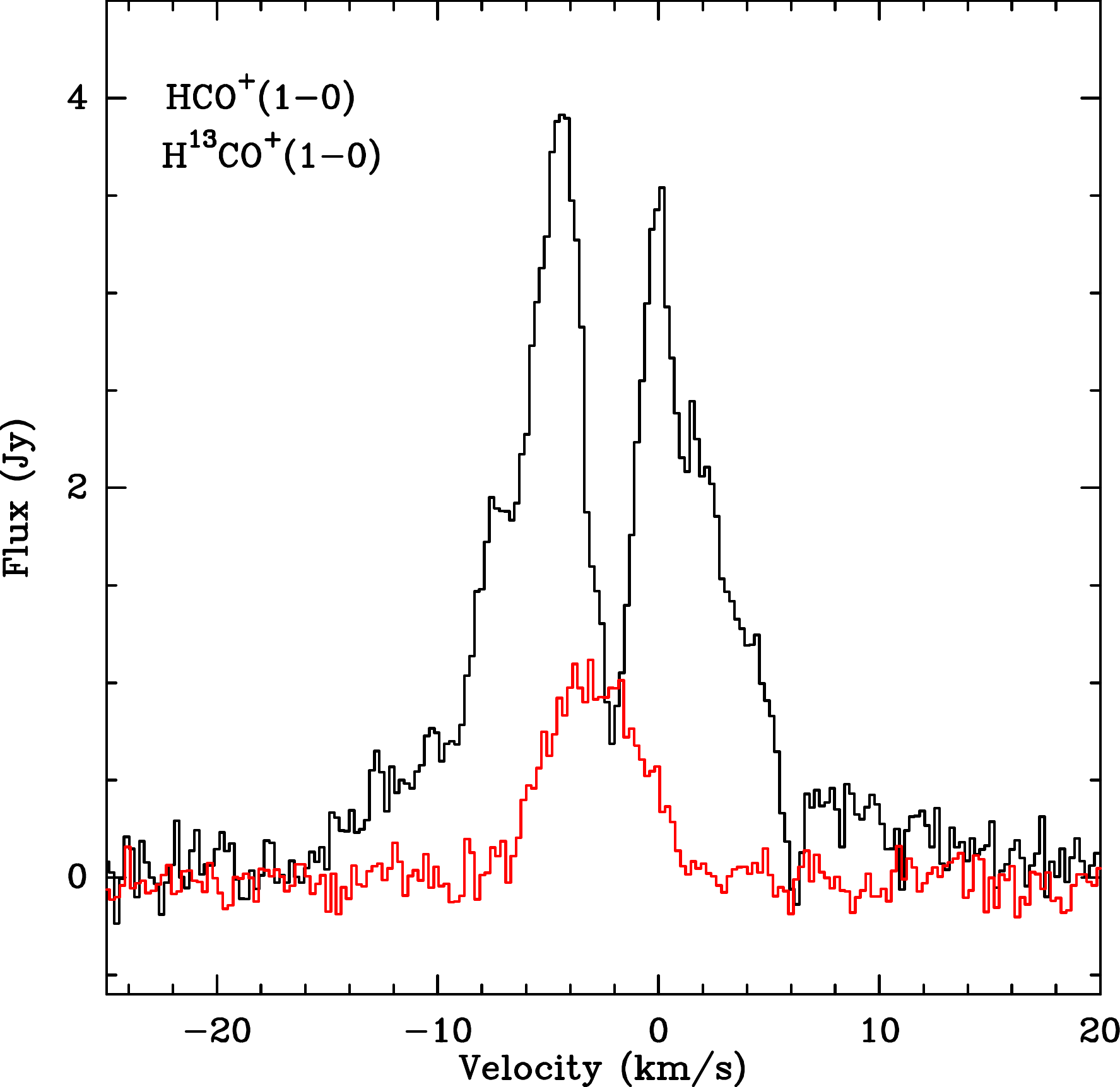}
\caption{MALT90 HCO$^+$\,(1--0) (black line) and H$^{13}$CO$^+$\,(1--0) (red line) spectra of Clump-1 integrated over a radius of $\sim0.3$\,pc.}
\label{malt90}
\end{figure}

The critical mass
  ratio, $M_{\rm line}/M_{\rm line, vir}$, of G351 is 0.7--0.9.
 Given the large
uncertainties in the mass estimate, we suggest that G351 is globally
in a quasi-stable equilibrium supported by a combination of thermal and
turbulent motions. Locally, the cloud may be unstable and collapse.
Indeed, we checked the MALT90 \citep{2011ApJS..197...25F,2013PASA...30...38F,2013PASA...30...57J} HCO$^+$, and H$^{13}$CO$^+$\,(1--0) spectra of Clump-1
to look for infall. While the spectra extracted at the position of the dust peak do not show any blue-skewed profiles
 \citep[see Fig.\,2 of][]{2016PASA...33...30R}, the data integrated over a radius of $\sim0.3$\,pc are suggestive of infall on those scales (see Fig.\,\ref{malt90}). However, the blue asymmetry in the HCO$^+$\,(1--0) disappear on larger scales, indicating local gravitational collapse.
This is similar to the
findings of \citet{2012ApJ...756L..13H} who found clear signs of
star formation activity, through the presence mid-IR point sources, in
gravitationally stable structures.

\section{Conclusions}

In this Paper, we have investigated the temperature and mass
distribution of G351.776--0.527 together with the kinematics of
moderately dense gas in the region using dust thermal continuum
emission maps from APEX and {\it Herschel} at different wavelengths
and APEX (C$^{18}$O\,(2--1))
spectroscopic data.  Thanks to the proximity of the source, even the
coarse angular resolution of single dish observations allows us to
construct a detailed picture of the region.  The source consists of a
main filamentary structure with a constant width of $\sim 0.2$\,pc and
a length of $\sim 4.6$\,pc surrounded by a complex network of branches
seen in absorption at mid-IR wavelengths and in emission in the
FIR. The main filament and network of branches are well
separated in H$_2$ column density; there is an average value in the
skeleton of G351 of $8.2\times10^{22}$\,cm$^{-2}$ and a maximum of
$1.5\times10^{22}$\,cm$^{-2}$ in the network. The most massive dust
condensation in G351 seems to divide the main filament in a cold
(11\,K--13\,K) quiescent northern part and a more active warmer
(27\,K--40\,K) southern part. The whole region has a mass of at least
$\sim2600$\,M$_\odot$, 20\% of which is in the
branches. Unfortunately, our current dataset does not allow us to
understand whether or not there is accretion of material from the branches
into G351.

The velocity dispersion of C$^{18}$O is relatively constant along the
main filament on scales of $\sim 0.2$\,pc with a typical value of
$\sim1$\,km\,s$^{-1}$.  C$^{18}$O is well fit with a single Gaussian
component along the spine of G351, but it shows two velocity
components in the majority of positions in the low-density  gas surrounding  G351,  and
in the branches. Under the assumption that G351 is a
single structure, its critical mass ratio is 0.7--0.9. We suggest
that G351 is globally in a quasi-stable situation supported by a
combination of thermal and turbulent motions close to virial
equilibrium, but that locally the filament has become unstable and
star formation has started. Finally, we detect velocity fluctuations
along the spine of G351 C$^{18}$O  in at least
one of the branches of the network in C$^{18}$O. We also find  local velocity gradients close to the dust peaks on top of a large-scale velocity gradient.

\section*{Acknowledgements}
We thank the anonymous referee for her/his accurate review,
which significantly contributed to improving the quality
of this paper.
We thank Axel Weiss for help with the SABOCA data reduction. E.S. acknowledges financial support from the VIALACTEA Project, a Collaborative Project under Framework Programme 7 of the European Union, funded under Contract \#607380. This work was partly supported by the Italian Ministero dell'Istruzione, Universit\`a e Ricerca through the grant Progetti Premiali 2012 -- iALMA (CUP C52I13000140001), and by the DFG cluster of excellence Origin and Structure of the Universe (\href{http://www.universe-cluster.de}{www.universe-cluster.de}). This work was partially carried out within the Collaborative Research Council 956, subproject A6, funded by the Deutsche Forschungsgemeinschaft (DFG). This  research
made  use  of  APLpy,  an  open-source  plotting  package  for  Python \citep{2012ascl.soft08017R}, of Astropy\footnote{http://www.astropy.org}, a community-developed core Python package for Astronomy \citep{astropy:2013, astropy:2018}, and of Matplotlib \citep{Hunter:2007}.




\begin{appendix}
  \section{Additional material}\label{app_fig}

  \begin{figure*}
\centering
\subfigure[][]{\includegraphics[width=0.42\textwidth]{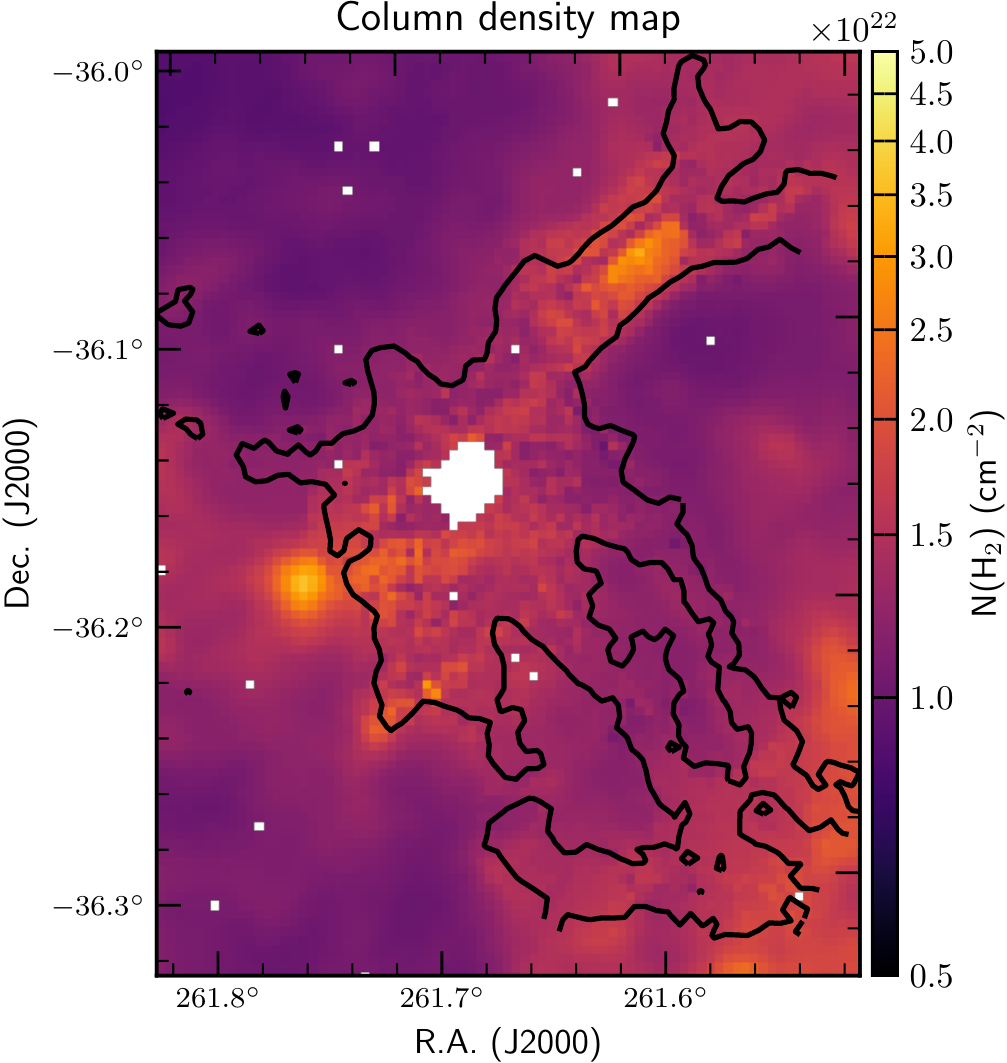}}
\subfigure[][]{\includegraphics[width=0.42\textwidth]{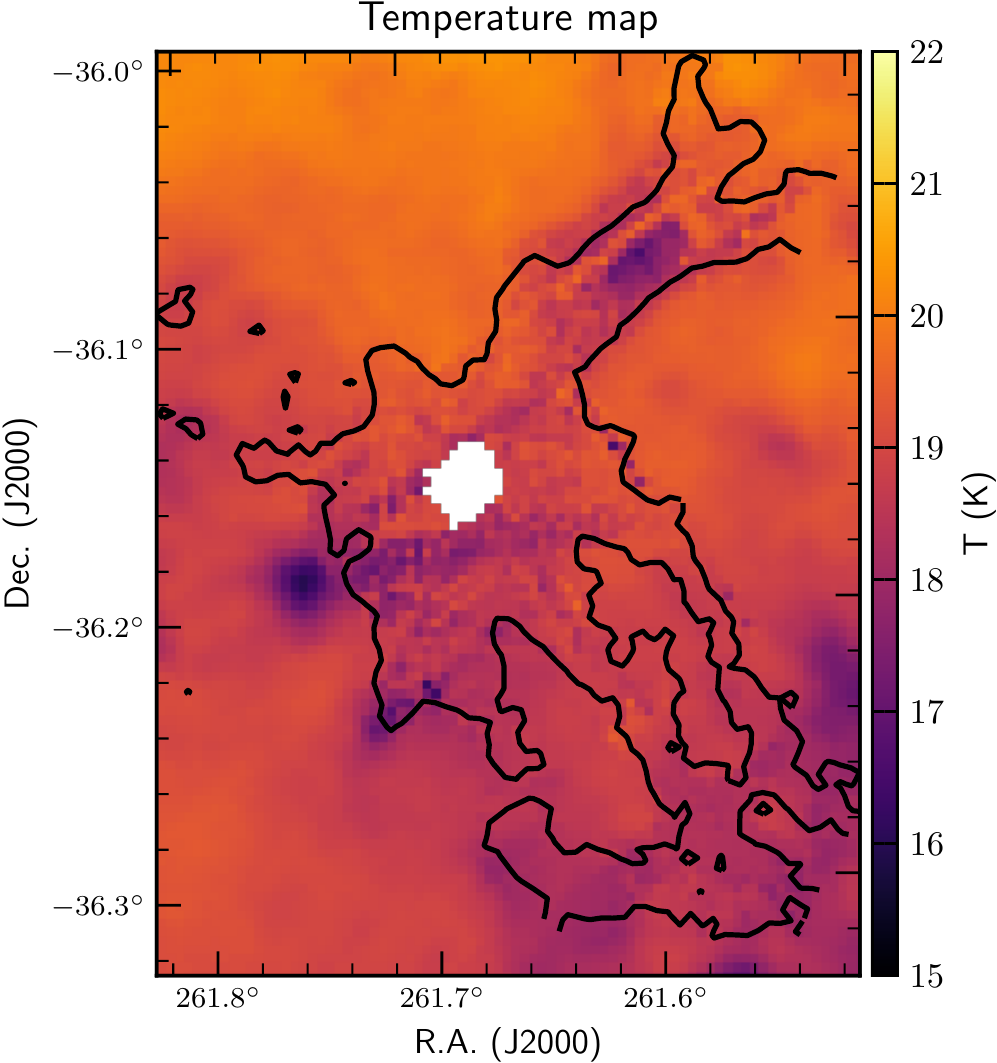}}
\caption{Distribution of the $N_{\rm{H_2}}$ column density (left panel) and dust temperature (right panel) for the background component. The solid black contour indicates the region in which the filamentary network component is defined.}
\label{bkgrd}
\end{figure*}

 \begin{figure*}[!htb]
\centering
\includegraphics[width=0.9\textwidth]{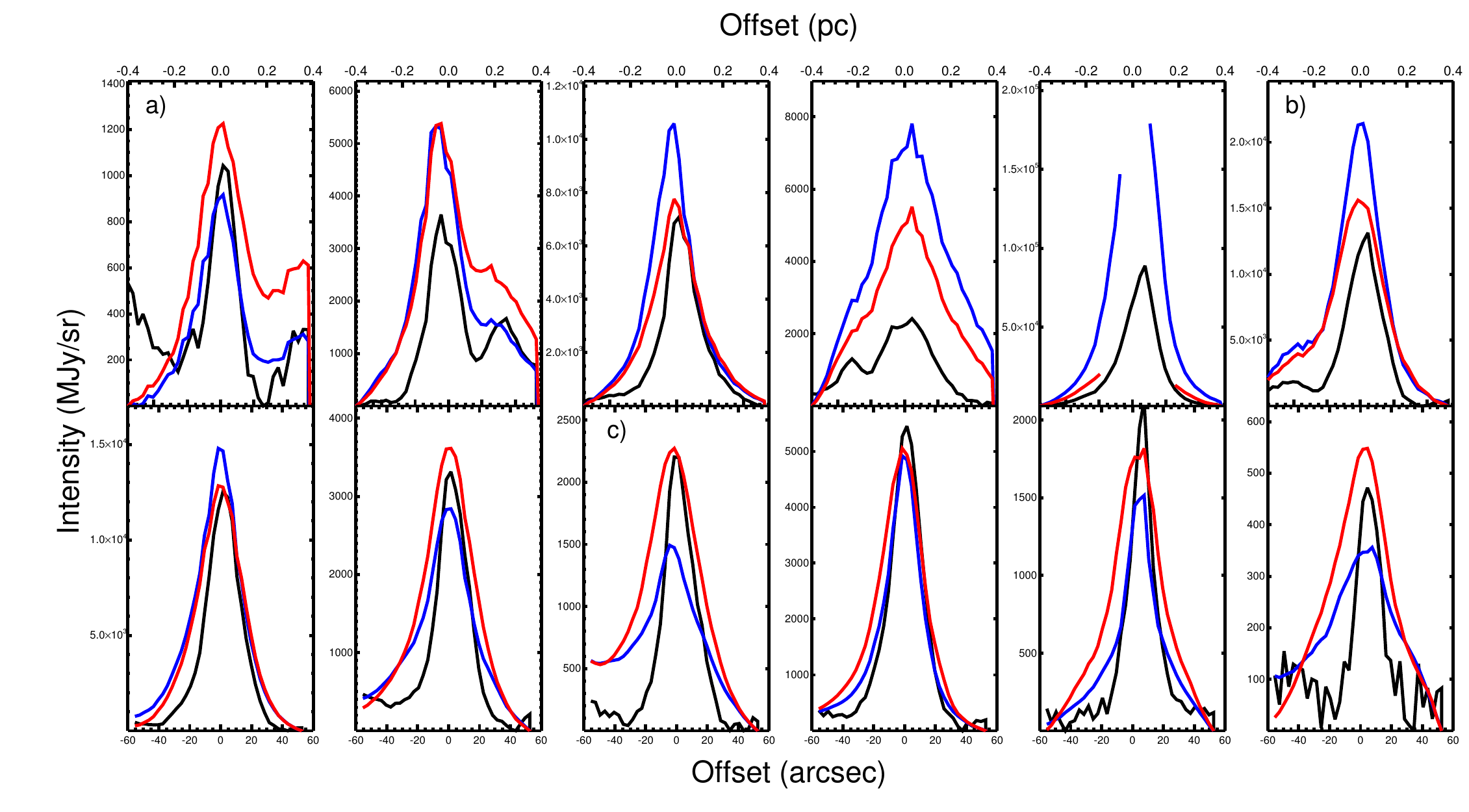}
\caption{Average radial profile of G351 estimated from the SABOCA emission at 350\,$\mu$m emission
  (solid black line),  Hi-GAL-250\,$\mu$m (red), and Hi-GAL-160\,$\mu$m data (blue). The panels show the 12 regions of
  Fig.\,\ref{saboca} ordered from top left to bottom right.}
\label{saboca_spire}
\end{figure*}

\begin{figure}[!htb]
\centering
\includegraphics[width=0.9\columnwidth]{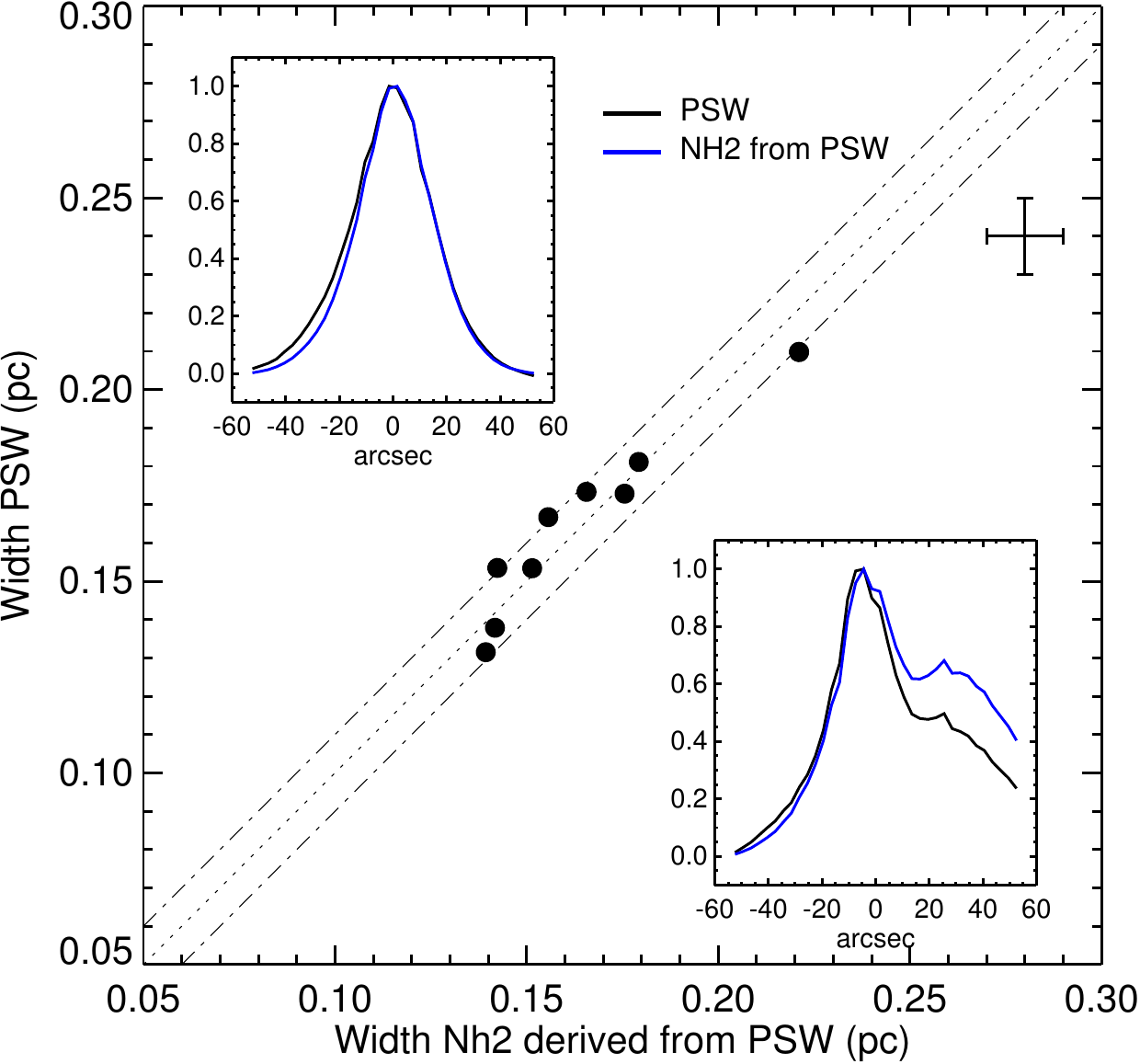}
\caption{Comparison between the widths of the 12 regions described in Sect.\,\ref{sec_g351} computed on the 250\,$\mu$m photometric map ($y-$axis) with those derived on the column density map obtained from the 250\,$\mu$m Hi-GAL data ($x-$axis). The widths are not convolved for the beam of the data. In the two panels, the averaged radial profiles for the second  (bottom right) of Fig.\,\ref{saboca}  and  seventh regions (top left) from the bottom  are shown derived with both methods.}
\label{temp}
\end{figure}

\begin{figure*}
\includegraphics[width=0.9\textwidth]{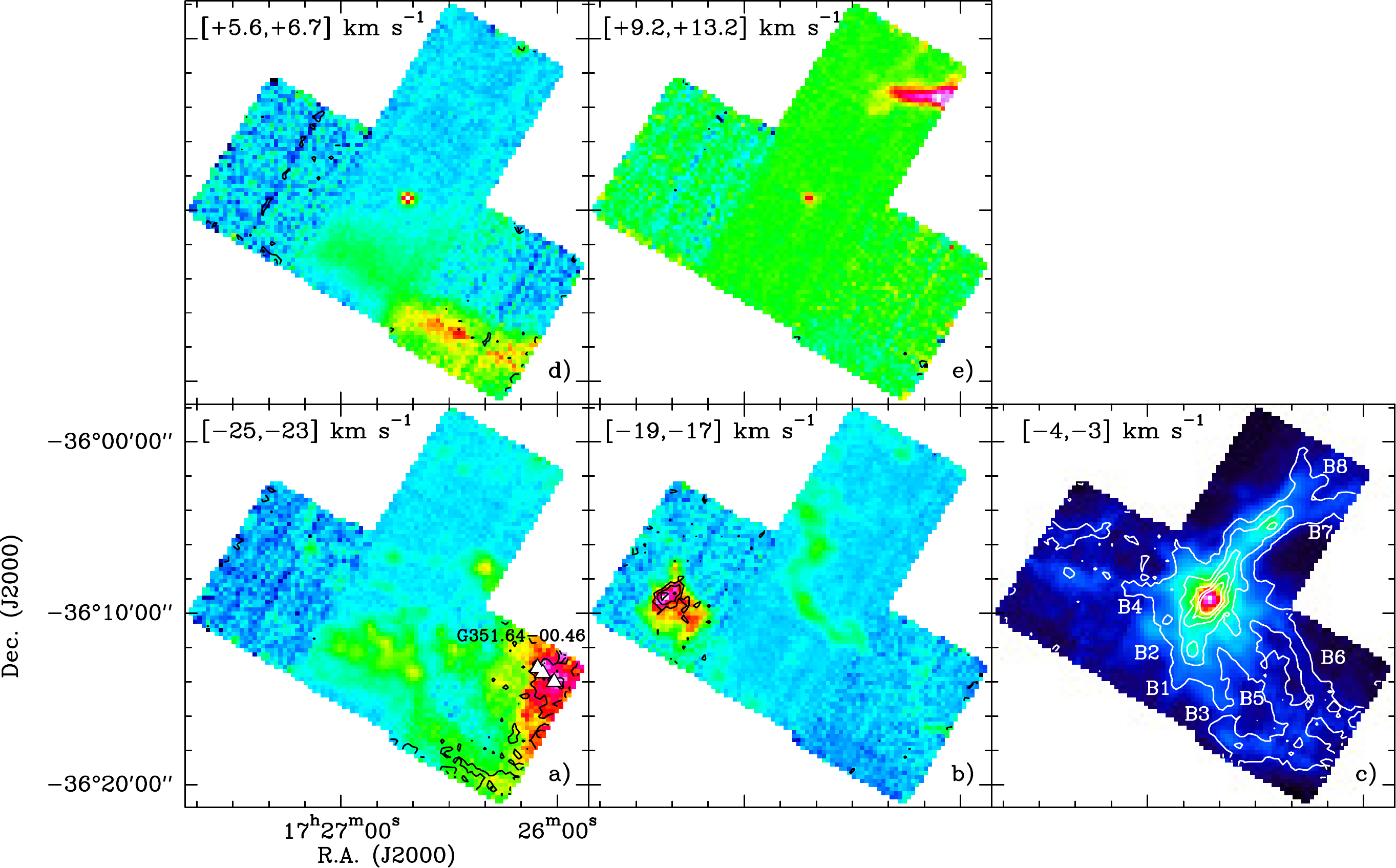}
\caption{Integrated emission of the $^{13}$CO\,(2--1) line over five different velocity ranges (see Fig.\,\ref{spectra}) is shown in colour scale.  The point-like emission seen towards Clump-1 in the $[+5.6,+6.7]$\,km\,s$^{-1}$ and $[+9.2,+13.2]$\,km\,s$^{-1}$ velocity ranges is due to high-velocity emission associated with molecular outflows \citep[see][]{2009A&A...507.1443L}.
  Black contours show the corresponding integrated C$^{18}$O\,(2--1) emission from 30\% of its peak emission in steps of 30\%. The white contours in the bottom right panel show the  integrated C$^{18}$O\,(2--1) emission in the range $-3.5\pm0.5$\,km\,s$^{-1}$ from 5\% to 45\% of the corresponding peak emission in steps of 10\%. The B1--B8 labels are as in Fig.\,\ref{fig1}.}
\label{clouds}
\end{figure*}

\begin{figure*}
\centering\includegraphics[width=0.85\textwidth]{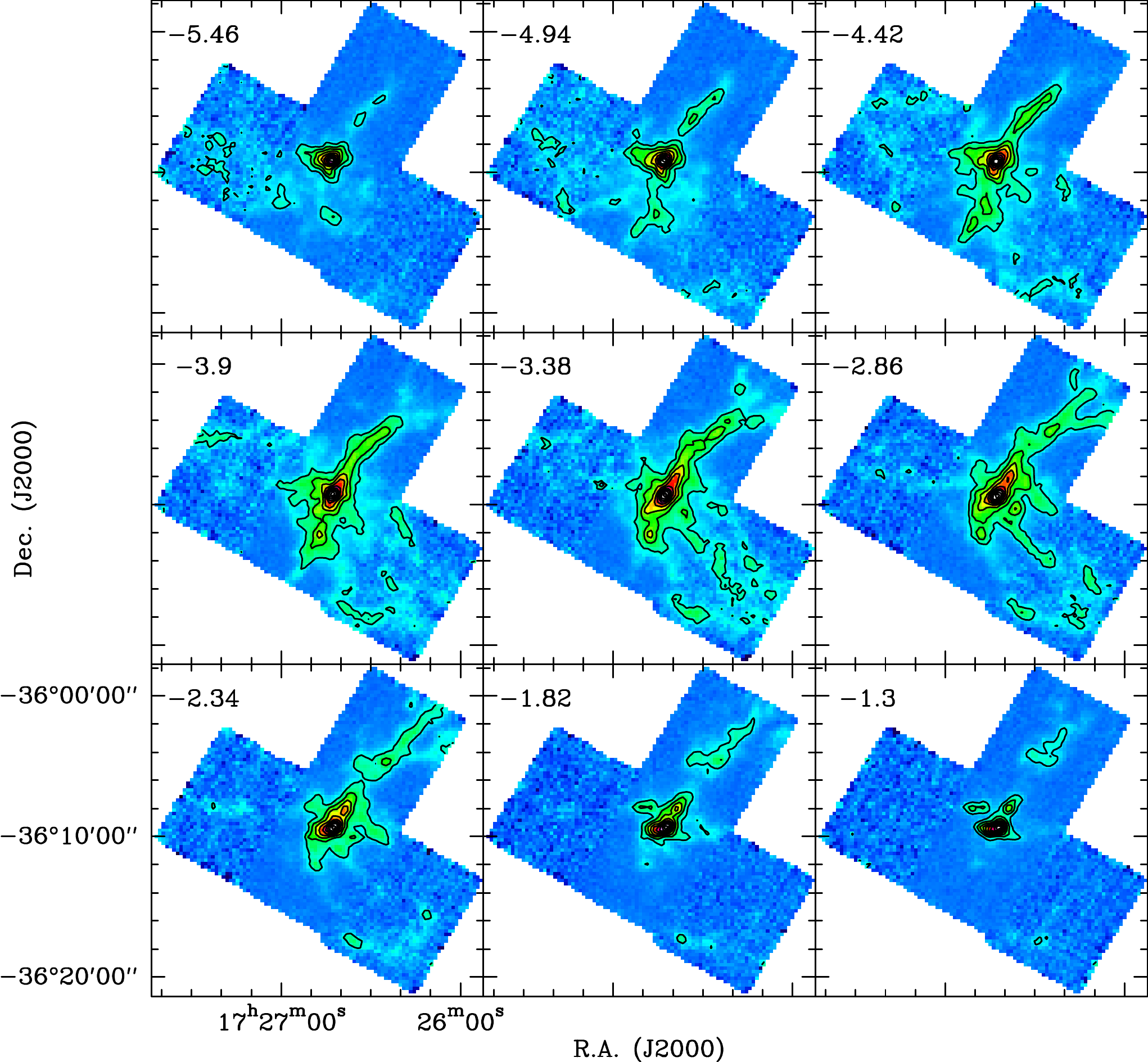}
\caption{Channel maps of C$^{18}$O\,(2--1). Black contours represent from 10\% of the peak intensity of each channel in steps of  10\%. The data were smoothed to a spectral resolution of $\sim 0.5$\,km\,s$^{-1}$.}
\label{chanmap}
\end{figure*}

\begin{figure}
\centering
\centering\includegraphics[width=0.85\columnwidth,angle=-90]{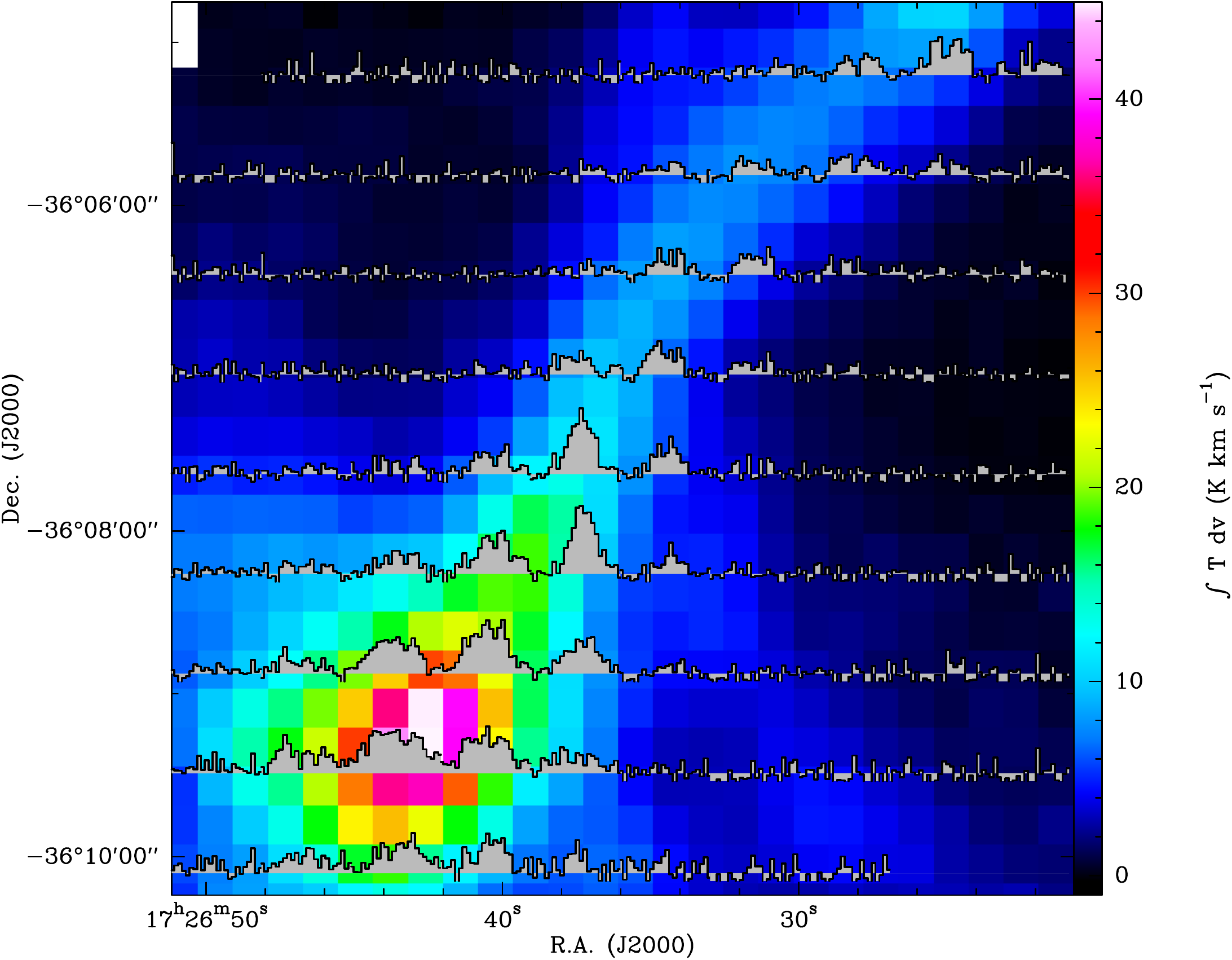}
\caption{Map of the N$_2$H$^+$\,(1--0) transition overlaid on the integrated emission of the same line  in the velocity range  $[-6,-1]$\,km\,s$^{-1}$. }
\label{n2hp_spectra}
\end{figure}

\begin{figure*}
\centering
\includegraphics[width=0.97\textwidth]{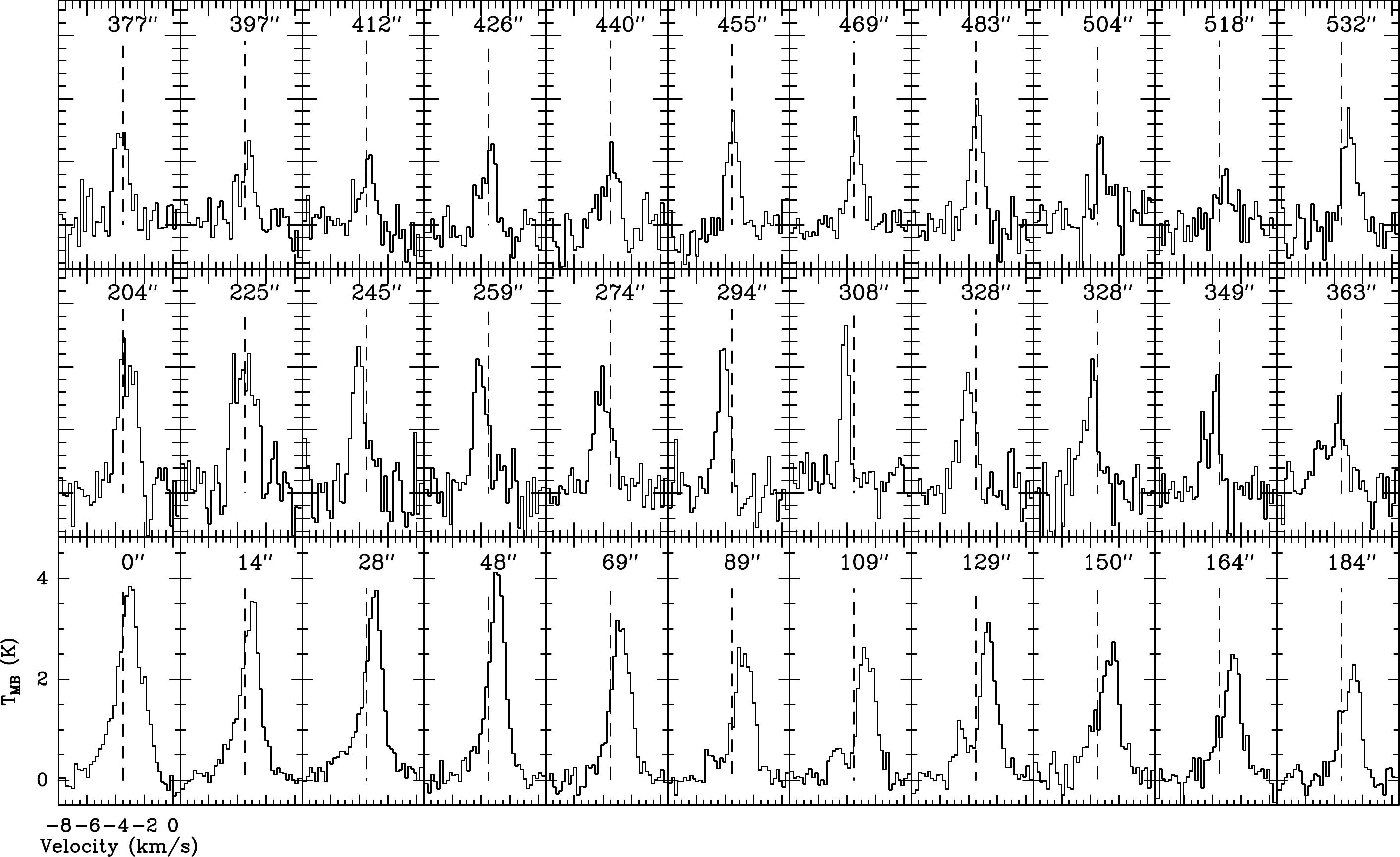}
\caption{C$^{18}$O spectra along Branch-6. The dashed line indicates a reference velocity of $-3.5$\,km\,s$^{-1}$ (see Fig.\,\ref{spectra}).
  The numbers in each panel indicate the angular distance from the spine of G351 along the branch.}
\label{br7}
\end{figure*}

 In this section, we provide additional material.
 Figure\,\ref{bkgrd}
 shows the distribution of the column density and dust temperature for the background component used in the two emission component fit of the \textit{Herschel} fluxes. Figure\,\ref{saboca_spire} presents the average radial profile of the G351 filament extracted from the photometric images at 350\,$\mu$m from SABOCA, and at 250\,$\mu$m and 160\,$\mu$m from Hi-GAL to illustrate how the width of G351 varies along its length. An estimate of the uncertainties introduced in our analysis by our choice of estimating the width of G351 using the photometric images is given in Fig.\,\ref{temp}. Additional information on the velocity field of the region is given in Figs.\,\ref{clouds}--\ref{br7}.

\end{appendix}

\end{document}